\documentclass[pra,twocolumn,aps,10pt]{revtex4-1}

\usepackage{amsmath,amssymb}

\usepackage{graphicx,color}

\usepackage{amsthm}

\usepackage[colorlinks=true]{hyperref}

\usepackage[T1]{fontenc}

\usepackage{bbm}

\newcommand{\ii}{\mathrm{i}}
\renewcommand{\vec}[1]{\mathbf{#1}}

\begin{document}

\title{Universal driving protocol for symmetry-protected Floquet topological phases}

\author{Bastian H{\"o}ckendorf}
\author{Andreas Alvermann}
\email{alvermann@physik.uni-greifswald.de}
\author{Holger Fehske}
\affiliation{Institut f{\"u}r Physik, Universit{\"a}t Greifswald, 17487 Greifswald, Germany}

\begin{abstract}
We propose a universal driving protocol for the realization of symmetry-protected topological phases in $2+1$ dimensional Floquet systems.
Our proposal is based on the theoretical analysis of the possible symmetries of a square lattice model with pairwise nearest-neighbor coupling terms.
Among the eight possible symmetry operators we identify the two relevant choices for topological phases with either time-reversal, chiral, or particle-hole symmetry.
From the corresponding symmetry conditions on the protocol parameters, we obtain the universal driving protocol where each of the symmetries can be realized or broken individually.
We provide specific parameter values for the different cases,
and demonstrate the existence of symmetry-protected copropagating and counterpropagating topological boundary states. The driving protocol especially allows us to switch between bosonic and fermionic time-reversal symmetry, and thus between a trivial and non-trivial symmetry-protected topological phase,
through continuous variation of a parameter.
\end{abstract}

\maketitle

\section{Introduction}

Topological phases have become a central topic of condensed matter research over the last few decades~\cite{PhysRevLett.45.494, PhysRevLett.49.405, PhysRevLett.95.146802, RevModPhys.82.3045, RevModPhys.82.3045, Konig766, PhysRevLett.98.106803}.
Recently, topological phases in periodically driven systems~\cite{PhysRevB.82.235114, Floquettopological, PhysRevB.84.235108, Flaschner,Wang453, PhysRevB.91.241404, Wang_SR} have attracted increasing interest,
including the anomalous Floquet topological insulators that exhibit a non-trivial topological phase although each of the individual Floquet bands is topologically trivial~\cite{PhysRevX.3.031005}.
Photonic lattices of evanescently coupled waveguides are especially well suited for the realization of these new topological phases~\cite{nature12066,RevModPhys.91.015006}.
In photonic lattices, periodic driving is replaced by spatially periodic modulation of the inter-waveguide distance, and thus of the coupling between adjacent waveguides,
such that one spatial coordinate represents the time axis of a $2+1$ dimensional Floquet system~\cite{nphoton.2014.248, 0953-4075-43-16-163001}.
In this way, direct implementation of driving protocols for (anomalous) Floquet topological insulators becomes possible~\cite{anomalous, anomalous_2}.

In this paper, we propose a universal driving protocol for symmetry-protected Floquet topological phases.
Presently, most of the theoretical proposals for the realization of such phases~\cite{PhysRevLett.112.026805, PhysRevB.90.195419, PhysRevB.90.205108, Zhou2014, PhysRevLett.114.106806, 1367-2630-17-12-125014, PhysRevB.93.075405,PhysRevB.96.155118,PhysRevB.96.195303, driven_KM, PhysRevB.93.115429, HAF18} focus on solid state applications and utilize mechanisms that are not well-suited for a photonic lattice implementation, involving, e.g., spin degrees of freedom~\cite{PhysRevLett.114.106806, driven_KM}, complicated driving schemes~\cite{1367-2630-17-12-125014, PhysRevB.93.075405}, or complex gauge potentials~\cite{PhysRevLett.112.026805, PhysRevB.90.195419, PhysRevB.90.205108, Zhou2014, PhysRevB.93.075405}.
The driving protocol proposed here, in contrast, has minimal complexity:
With only six steps per period and simple pairwise couplings between adjacent sites of a square lattice it can realize
Floquet topological phases with  time-reversal, chiral, or particle-hole symmetry.
Given its minimal complexity, the protocol is not only of intrinsic theoretical value,
but allows for immediate experimental observation of these symmetry-protected topological phases in photonic systems.

The starting point for our construction is the analysis of
the possible symmetry operators 
for a driving protocol with only pairwise couplings.
On a square lattice, eight distinct symmetry operators have to be considered,
but only two of them can lead to driving protocols with symmetry-protected topological phases. 

The symmetry analysis provides us with the general form of the driving protocol, 
which appears in two types:
a protocol \textsf{A} that supports time-reversal symmetry,
and a protocol \textsf{B} that supports particle-hole symmetry.
These two types of the universal driving protocol cover all four symmetry combinations 
with non-trivial $2+1$-dimensional topological phases.
To verify the universality of the protocol we provide specific parameter sets according to the conditions enforced by the different symmetries.
The symmetry-protected Floquet topological phases realized with these parameters are analyzed by means of symmetry-adapted topological bulk invariants~\cite{PhysRevLett.114.106806,1367-2630-17-12-125014,HAF18}, and transport via counterpropagating boundary states is demonstrated numerically.

The structure of the paper is as follows: In Sec.~\ref{Sec_2} we define the square lattice model that is the basis of the present study.
In Sec.~\ref{sec:symm} we identify the eight types of symmetry operators that are compatible with the assumptions made in the construction of the square lattice model,
analyze which of these operators can be used to implement time-reversal, chiral, or particle-hole symmetry,
and determine the resulting constraints on the driving protocol. 
The universal driving protocol is then introduced and investigated in Secs.~\ref{sec:principal}--\ref{sec:PH},
once with a focus on time-reversal symmetry (protocol \textsf{A} in Secs.~\ref{Sec_4}, \ref{Sec_5}), once for particle-hole symmetry (protocol \textsf{B} in Sec.~\ref{sec:PH}). We conclude in Sec.~\ref{Sec_6}.
The appendices detail the pseudo-spin interpretation of our construction (App.~\ref{app:Pseudo}),
explore the differences between  parallel and antiparallel diagonal couplings (Apps.~\ref{app:Parallel},~\ref{app:Combined}), and explain why a universal driving protocol with a two-site unit cell should not exist (App.~\ref{app:nogo}).

\section{Square lattice model}
\label{Sec_2}

\begin{figure*}
\hspace*{\fill}
\includegraphics[scale=0.9]{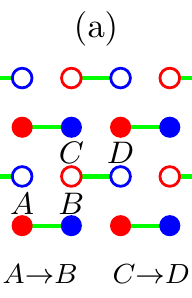}
\hspace*{\fill}
\includegraphics[scale=0.9]{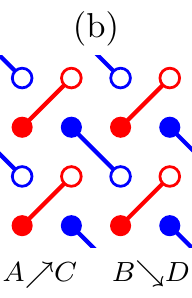}
\hspace*{\fill}
\includegraphics[scale=0.9]{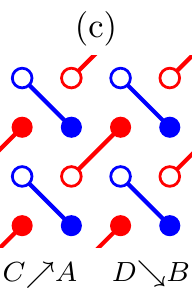}
\hspace*{\fill}
\includegraphics[scale=0.9]{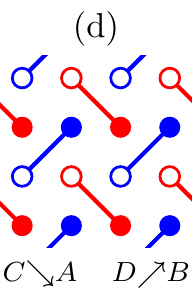}
\hspace*{\fill}
\includegraphics[scale=0.9]{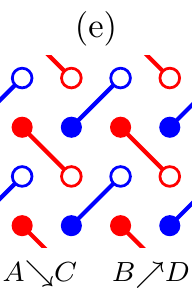}
\hspace*{\fill}
\includegraphics[scale=0.9]{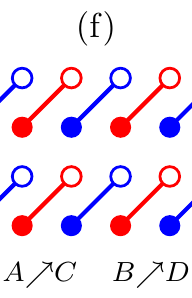}
\hspace*{\fill}
\includegraphics[scale=0.9]{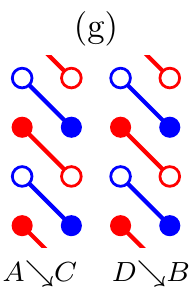}
\hspace*{\fill}
\includegraphics[scale=0.9]{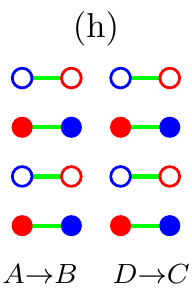}
\hspace*{\fill}
\caption{Eight examples of patterns of compatible pairwise couplings in the square lattice model.
The couplings are indicated by lines between adjacent sites.
The labels below each pattern follow the notation in the text, with the four sites $A$ (filled red), $B$ (filled blue), $C$ (open red), $D$ (open blue) as shown in the leftmost pattern.
 }
\label{fig:Lattice}
\end{figure*}

For the construction of the driving protocol we start with a square lattice with a four-element unit cell,
which is the minimal choice for the realization of topological phases with, e.g., time-reversal symmetry (see App.~\ref{app:nogo} for the case of a two-element unit cell).
The lattice sites are located at positions $\vec r = i \vec a_x + j \vec a_y + \boldsymbol \delta_s$, with $i,j \in \mathbb Z$ and $s \in \{A,B,C,D \}$.
Here, $\vec a_{x}=(2,0)^t, \vec a_y=(0,2)^t$ are the primitive vectors of lattice translations,
and $\boldsymbol \delta_A = (0,0)^t$,
$\boldsymbol \delta_B = (1,0)^t$,
$\boldsymbol \delta_C = (1,1)^t$,
$\boldsymbol \delta_D = (2,1)^t$
enumerate the four sites in the unit cell  (see Fig.~\ref{fig:Lattice}).
This enumeration is purely a matter of convention, but the present choice will prove useful later.
All vectors are measured as multiples of some unspecified unit of length.

On the square lattice, pairwise coupling of neighboring lattice sites
can occur along four directions: horizontal ($\boldsymbol \delta_{\rightarrow}=(1,0)$),
vertical ($\boldsymbol \delta_{\uparrow}=(0,1)$),
diagonal ($\boldsymbol \delta_{\nearrow}=(1,1)$),
and anti-diagonal ($\boldsymbol \delta_{\searrow}=(1,-1)$).
This gives $4 \times 4 = 16$ translational invariant pairwise coupling terms 
$\hat t_{s \circ s'} = \sum_{\vec r =  i \vec a_x + j \vec a_y} |\vec r + \boldsymbol \delta_s + \boldsymbol \delta_\circ \rangle \langle \vec r + \boldsymbol \delta_s |$,
with $s \in \{A, B, C, D \}$ and $\circ \in \{\searrow, \rightarrow, \nearrow, \uparrow \} $.
In essence, $\hat t_{s \circ s'}$ moves a particle (representing, e.g., light in a waveguide) from sites of type $s$ along direction $\circ$ to sites of type $s'$.
The Hermitian conjugate $\hat t_{s \circ s'}^\dagger$ operates in the opposite direction, from $s'$ to $s$.
Note that $s'$ is determined by $s$ and $\circ$, and included for notational clarity only.
In addition to the pairwise coupling terms, there are four on-site terms
$\hat n_s = \sum_{\vec r =  i \vec a_x + j \vec a_y} |\vec r + \boldsymbol \delta_s \rangle \langle \vec r + \boldsymbol \delta_s |$, which involve a single type $s$ of lattice sites.

Note that we do not use the language of second quantization but the simpler bra-ket notation.
In particular, we do not fix the particle statistics, as encoded by the (anti-) commutation relations of creation and annihilation operators in second quantization, and consider both fermionic and bosonic symmetries for the driving protocol.

\begin{figure*}[t]
\hspace*{\fill}
\includegraphics[scale=1]{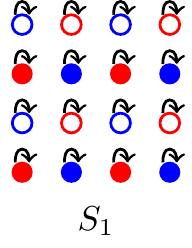}
\hspace*{\fill}
\includegraphics[scale=1]{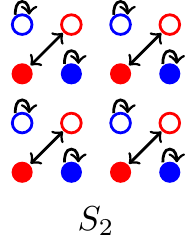}
\hspace*{\fill}
\includegraphics[scale=1]{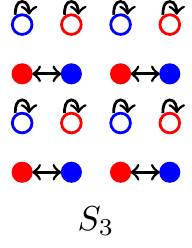}
\hspace*{\fill}
\includegraphics[scale=1]{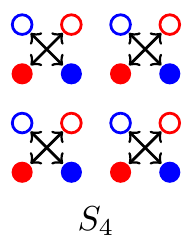}
\hspace*{\fill}
\includegraphics[scale=1]{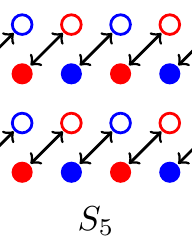}
\hspace*{\fill}
\includegraphics[scale=1]{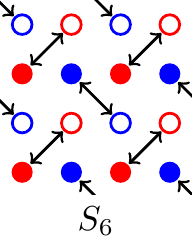}
\hspace*{\fill}
\includegraphics[scale=1]{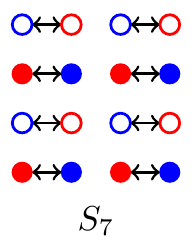}
\hspace*{\fill}
\includegraphics[scale=1]{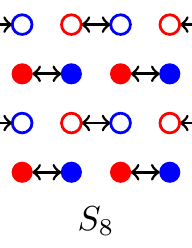}
\hspace*{\fill}
\\
\hspace*{\fill}
\includegraphics[scale=1]{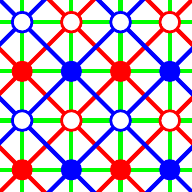}
\hspace*{\fill}
\includegraphics[scale=1]{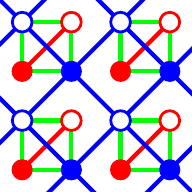}
\hspace*{\fill}
\includegraphics[scale=1]{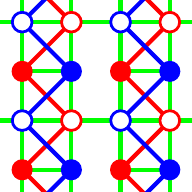}
\hspace*{\fill}
\includegraphics[scale=1]{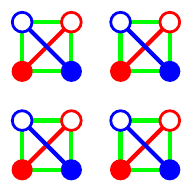}
\hspace*{\fill}
\includegraphics[scale=1]{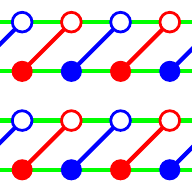}
\hspace*{\fill}
\includegraphics[scale=1]{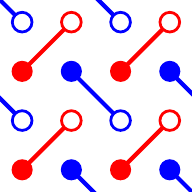}
\hspace*{\fill}
\includegraphics[scale=1]{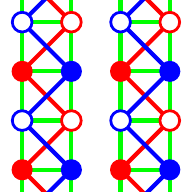}
\hspace*{\fill}
\includegraphics[scale=1]{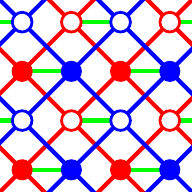}
\hspace*{\fill}
\caption{Graphical representation of the eight symmetry operators $S_1, \dots, S_8$ for the Hamiltonian~\eqref{Ham}. 
The arrows in the upper row indicate how each symmetry operator maps a site onto itself, or onto one of its eight neighbors.
The lines in the lower row indicate the compatible pairwise coupling terms.
}
\label{fig:Symm_op}
\end{figure*}

The general square lattice Hamiltonian reads
\begin{equation}\label{Ham}
\begin{split}
 H(t) =  & \sum_{\substack{s \in \{A,B,C,D\} \\ \circ \in \{\searrow, \rightarrow, \nearrow, \uparrow \}  } }   J_{s \circ s'}(t)  \, \hat t_{s \circ s'} +  J_{s \circ s'}(t)^*  \, \hat t_{s \circ s'}^\dagger \\ + & \sum_{s \in \{A,B,C,D\}} \Delta_s(t) \,\hat n_s \;.
 \end{split}
\end{equation}
It includes $4 \times 4 + 4 = 20$ time-dependent parameters $J_{s \circ s'}(t)$ (for pairwise couplings) and $\Delta_s(t)$ (for on-site potentials). All parameters, and so $H(t)$ itself, will be periodic in time, with period $T$.

In the Hamiltonian~\eqref{Ham}, not all parameter combinations are admissible.
Instead, we impose a compatibility constraint on the pairwise couplings: terms that involve the same lattice site cannot occur together at the same time.
Therefore, $J_{s \circ s'}(t) \ne 0$ requires $J_{p \bullet p'}(t) = 0$ for any other coupling with $\{ p, p' \} \cap \{s,s'\} \ne \emptyset$.
Note that this compatibility condition is fulfilled precisely if the two operators $\hat t_{s \circ s'}$, $\hat t_{p \bullet p'}$ commute.
A further restriction concerns pairwise couplings that ``cross each other'' on the lattice, which occurs only for diagonal couplings. For example, $J_{A \nearrow C}(t)$, $J_{D \searrow B}(t)$ cannot both be non-zero at the same $t$.
The on-site potentials $\Delta_s(t)$ are not restricted, and can occur together with any pairwise coupling.

The combination of compatible pairwise couplings gives a total of $12$ diagonal $+$ $4$ horizontal + $4$ vertical = $20$ coupling patterns, eight of which are depicted in Fig.~\ref{fig:Lattice}.
For every coupling pattern, at most two parameters $J_{s \circ s'}(t)$ of the Hamiltonian are non-zero.
The driving protocol will consist of a cyclic sequence of these coupling patterns,
which are selected according to the symmetry analysis in the next section.

In the introduction, we have motivated the universal driving protocol also with the possibility of a photonic lattice implementation.
In such an implementation, lattice sites correspond to waveguides.
Since coupling of waveguides is achieved by reducing their distance locally ~\cite{0953-4075-43-16-163001},
spatially complex coupling patterns are not easily realized experimentally and thus should be avoided in the driving protocol.
This includes the coupling of more than two or of non-adjacent waveguides, and results in the constraints imposed on the Hamiltonian above.

\section{Symmetry operators and symmetry conditions}
\label{sec:symm}

The symmetry relations of time-reversal, chiral, and particle-hole symmetry, as specified further below,
involve a transformation $S H(t) S^{-1}$ of the Hamiltonian with a translational-invariant operator $S$.
The symmetry relations can only hold if the transformed Hamiltonian has the same structure as the original Hamiltonian, and is again composed only of
pairwise couplings and on-site potentials.

This observation 
restricts the possible symmetry operators $S$ in a similar way to the coupling patterns in Fig.~\ref{fig:Lattice}.
In particular, every operator can only be composed of non-overlapping pairwise terms $\hat t_{s \circ s'}$ or on-site terms $\hat n_s$.
Otherwise, with overlapping terms, the transformed Hamiltonian $S H(t) S^{-1}$ would contain couplings between three or more lattice sites. We do not, however, have the restriction that pairwise terms cannot cross.

In total, there are the eight possible symmetry operators $S_1, \dots, S_8$ shown in Fig.~\ref{fig:Symm_op}, not counting rotations, reflections, or translations. These operators map every lattice site onto exactly one other lattice site, either the same (e.g., for $S_1$) or a different one (e.g., for $S_8$).

Each symmetry operator is compatible with the pairwise couplings shown in the lower half of Fig.~\ref{fig:Symm_op}.
Exactly these couplings are mapped again to couplings between adjacent sites in the transformation 
$S H(t) S^{-1}$
with the respective symmetry operator. 
The remaining pairwise couplings are mapped onto coupling terms that do not occur in the Hamiltonian, and must be excluded. 

Two observations are immediate.
First, if the graph spanned by the compatible pairwise couplings in Fig.~\ref{fig:Symm_op} is disconnected, such that propagation  is restricted to a lower-dimensional subset of the lattice, non-trivial $2+1$ dimensional topological phases cannot exist.
The symmetry operators $S_5$ and $S_7$ restrict propagation to quasi-one-dimensional stripes, $S_4$ and $S_6$ to finite regions.
Only the operators $S_1$, $S_2$, $S_3$, $S_8$ allow for propagation on the entire two-dimensional lattice.
Second, among these four operators, $S_1$, $S_2$, $S_3$ involve isolated on-site terms $\hat n_s$.
Such terms necessarily square to 
$(\xi \hat n_s) (\xi \hat n_s)^* = |\xi|^2 \hat n_s$ for any $\xi \in \mathbb C$, 
which is incompatible with non-unitary symmetries that require $S S^* = -1$
(e.g., fermionic time-reversal symmetry with $\Theta^2=-1$).
These two observations leave us with the operator $S_8$ for the construction of the universal driving protocol.
Note that $S_8$ is compatible with the symmetry operator $S_1$,
which will allow us to implement an additional particle-hole symmetry once the protocol has been constructed with $S_8$.

The unitary operators $S_1$ and $S_8$ can be specified by two $2 \times 2$ unitary matrices $\sigma$, $\tau$
in the form
\begin{equation}\label{S18}
\begin{split}
S_{1,8} &= \sigma_{AA} \, \hat n_A + \sigma_{BB} \, \hat n_B + \sigma_{BA} \, \hat t_{A \rightarrow B}  + \sigma_{AB} \, \hat t_{A \rightarrow B}^\dagger \\
&\, + \tau_{CC} \, \hat n_C + \tau_{DD} \, \hat n_D + \tau_{DC} \, \hat t_{C \rightarrow D}  + \tau_{CD} \, \hat t_{C \rightarrow D}^\dagger \;,
\end{split}
\end{equation}
where $\sigma$, $\tau$ have only diagonal (for $S_1$) or only off-diagonal (for $S_8$) entries.
The mnemonic form of this equation is
\begin{equation}
\text{``} S_{1,8} = \left(\begin{smallmatrix} |A\rangle \\ |B\rangle \end{smallmatrix}\right) \cdot \boldsymbol \sigma \left(\begin{smallmatrix} \langle A| \\ \langle B| \end{smallmatrix}\right)
+ \left(\begin{smallmatrix} |C\rangle \\ |D\rangle \end{smallmatrix}\right) \cdot \boldsymbol \tau \left(\begin{smallmatrix} \langle C| \\ \langle D| \end{smallmatrix}\right) \text{''} \;.
\end{equation}
This expression suggests a pseudo-spin interpretation of the ``red'' and ``blue'' sublattice structure depicted in Figs.~\ref{fig:Lattice},~\ref{fig:Symm_op}, which is detailed in App.~\ref{app:Pseudo}. For the following considerations, this interpretation is not needed.

The operator $S_8$ is compatible with ten pairwise coupling terms, as depicted in Fig.~\ref{fig:Symm_op},
and all diagonal terms $\hat n_s$.
These fourteen terms change according to Table~\ref{tab:transform} under a transformation with the operator $S_1$ or $S_8$.
These transformation rules, together with the symmetry conditions specified next, determine the constraints on the parameters of the Hamiltonian for the respective symmetry, and thus the structure of the driving protocol.

Inspection of Figs.~\ref{fig:Lattice},~\ref{fig:Symm_op} shows how a transformation with the symmetry operators $S_1$ and $S_8$ affects different coupling patterns.
The symmetry operator $S_1$ maps every pattern onto itself.
For horizontal couplings,
the symmetry operator $S_8$ leaves pattern (a) invariant but is not compatible with pattern (h).
The remaining two horizontal patterns not shown in Fig.~\ref{fig:Lattice}, as well as all four patterns with vertical couplings, are also incompatible with $S_8$.
For patterns with perpendicular diagonal couplings,
$S_8$ swaps patterns (b) $\leftrightarrow$ (e) and (c) $\leftrightarrow$ (d).
For patterns with parallel diagonal couplings,
$S_8$ leaves pattern (f) invariant while pattern (g) is mapped onto a different pattern with parallel couplings (cf. App.~\ref{app:Parallel}).
Note that this behavior concerns only  the geometric structure of the coupling patterns. For the mapping of the coupling parameters, Table~\ref{tab:transform} has to be consulted.

\begin{table}
\caption{Transformation $T \mapsto S_{1} T S_{1}^{-1}$ and $T \mapsto S_{8} T S_{8}^{-1}$ of the fourteen terms $T$ used in the construction of the driving protocol.}
\begin{ruledtabular}
\begin{tabular}{ccc}
$T$ & $S_1 T S_1^{-1}$ & $S_8 T S_8^{-1}$ \\\hline
 $\hat t_{A \rightarrow B}$ & $\sigma_{AA}^* \sigma_{BB} \, \hat t_{A \rightarrow B}$ & $\sigma_{BA}^* \sigma_{AB} \, \hat t_{A \rightarrow B}^{\dagger}$\\[1pt]
 $\hat t_{C \rightarrow D}$ & $\tau_{CC}^* \tau_{DD} \, \hat t_{C \rightarrow D}$ & $\tau_{DC}^* \tau_{CD} \, \hat t_{C \rightarrow D}^{\dagger}$\\[1pt]
  $\hat t_{A \nearrow C}$ & $\sigma_{AA}^* \tau_{CC} \, \hat t_{A \nearrow C}$ & $\sigma_{BA}^* \tau_{DC} \, \hat t_{B \nearrow D}$\\[1pt]
   $\hat t_{A \searrow C}$ & $\sigma_{AA}^*\tau_{CC} \, \hat t_{A \searrow C}$ & $\sigma_{BA}^* \tau_{DC} \, \hat t_{B \searrow D}$\\[1pt]
   $\hat t_{C \nearrow A}$ & $\sigma_{AA}\tau_{CC}^* \, \hat t_{C \nearrow A}$ & $\sigma_{BA} \tau_{DC}^*  \, \hat t_{D \nearrow B}$\\[1pt]
 $\hat t_{C \searrow A}$ & $\sigma_{AA}\tau_{CC}^* \, \hat t_{C \searrow A}$ & $\sigma_{BA} \tau_{DC}^*  \, \hat t_{D \searrow B}$\\[1pt]
  $\hat t_{B \nearrow D}$ & $\sigma_{BB}^*\tau_{DD} \, \hat t_{B \nearrow D}$ & $\sigma_{AB}^* \tau_{CD} \, \hat t_{A \nearrow C}$\\[1pt]
  $\hat t_{B \searrow D}$ & $\sigma_{BB}^*\tau_{DD} \, \hat t_{B \searrow D}$ & $\sigma_{AB}^* \tau_{CD} \, \hat t_{A \searrow C}$\\[1pt]
  $\hat t_{D \nearrow B}$ & $\sigma_{BB}\tau_{DD}^* \, \hat t_{D \nearrow B}$ & $\sigma_{AB} \tau_{CD}^* \, \hat t_{C \nearrow A}$\\[1pt]
  $\hat t_{D \searrow B}$ & $\sigma_{BB}\tau_{DD}^* \, \hat t_{D \searrow B}$ & $\sigma_{AB} \tau_{CD}^* \, \hat t_{C \searrow A}$\\[1pt]
  $\hat n_A$ & $\hat n_A$ & $\hat n_B$\\
  $\hat n_B$ & $\hat n_B$ & $\hat n_A$\\
  $\hat n_C$ & $\hat n_C$ & $\hat n_D$\\
  $\hat n_D$ & $\hat n_D$ & $\hat n_C$
\end{tabular}
\end{ruledtabular}
\label{tab:transform}
\end{table}

\subsection{Time-reversal symmetry}
\label{Sec_3TR}

The symmetry relation for time-reversal symmetry is
\begin{equation}
H_{\mathrm{tr}}(T-t)=\Theta H_{\mathrm{tr}}(t)\Theta^{-1}\; ,
\label{eq:tr_symm}
\end{equation} 
with an anti-unitary symmetry operator $\Theta$ for which $\Theta^2 = \pm 1$.
For our purposes, the operator $\Theta$ can be written in the form $\Theta = K S_8$, with the unitary symmetry operator $S_8$ from the previous section and complex conjugation $K$. Then, the condition $\Theta^2 = \pm 1$ is equivalent to
$\sigma^* \sigma = \tau^* \tau = \pm 1$. 

For fermionic time-reversal symmetry with $\Theta^2 = -1$, the only choice is $\sigma = \alpha \sigma_y$ and $\tau = \beta \sigma_y$, with the Pauli matrix $\sigma_y$ and two phases $\alpha, \beta \in \mathbb C$, $|\alpha| = |\beta| =1$.
Without loss of generality, we set $\alpha =  \beta = 1$ such that the transformation of (anti-)diagonal couplings in Table~\ref{tab:transform} involves the same sign.
 The relevant operator $S_8$ 
 thus is
\begin{equation}
S_8 = \ii (\hat t_{A \rightarrow B}  - \hat t_{A \rightarrow B}^\dagger) + \ii (\hat t_{C \rightarrow D} - \hat t_{C \rightarrow D}^\dagger ) \;,
\end{equation}
that is $\sigma_{AB} = - \sigma_{BA} = \tau_{CD} = - \tau_{DC} = - \ii$.
Note that the operator does not involve on-site terms $\hat n_s$.
The resulting conditions on the parameters of the Hamiltonian following from Eq.~\eqref{eq:tr_symm} are given in Table~\ref{tab:symm}.

For bosonic time-reversal symmetry with $\Theta^2 = 1$, we must have
$\sigma^* \sigma = \tau^* \tau = 1$,
and choose $\sigma = \tau = \sigma_x$ with the Pauli matrix $\sigma_x$.

\subsection{Chiral symmetry}
\label{Sec_3CH}

The symmetry relation for chiral symmetry is
\begin{equation}\label{eq:ch_symm}
 H_\mathrm{ch}(T-t) = -  \Gamma H_\mathrm{ch}(t) \Gamma^{-1}
\end{equation}
with a unitary operator $\Gamma$ and, by convention, $\Gamma^2=1$. 
Note that, in difference to unitarily-realized symmetries, this relation contains a minus sign:
the Hamiltonian anti-commutes with $\Gamma$.

In the universal driving protocol, which will be constructed based on the symmetry operator $S_8$,
chiral symmetry can be implemented either by means of $S_1$ or $S_8$. Here, we deliberately choose the operator $S_8$ because of its overall significance in the present constructions.

In order to obtain a symmetry-protected phase, chiral symmetry must be realized as a 
bipartite even-odd sublattice symmetry, where the operator $\Gamma$ includes a minus sign on every second unit cell~\cite{PhysRevB.93.075405, HAF18}.
With this alternating sign, we have
\begin{equation}
 \Gamma = \Big[ \sum_{\substack{\vec r =  i \vec a_1 + j \vec a_2 \\ s \in \{A,B,C,D\}}} (-1)^{i+j}  |\vec r + \boldsymbol \delta_s \rangle\langle \vec r + \boldsymbol \delta_s | \Big] S_8 
\end{equation}
as a modification of the translational-invariant operator $S_8$. 
The alternating sign depends on our choice of the unit cell of the square lattice, which here consists of the sites $A$, $B$, $C$, $D$ in Fig.~\ref{fig:Lattice}. This is the natural choice when dealing with the symmetry operator $S_8$.

The condition $\Gamma^2 =1$  is equivalent to $S_8^2=1$, that is $\sigma^2 = \tau^2 = 1$, since the alternating sign  cancels.
As for fermionic time-reversal symmetry, we choose $\sigma = \tau = \sigma_y$ with the Pauli matrix $\sigma_y$.
Note that this choice gives $\Gamma^2=1$ here,
 but $\Theta^2=-1$ for time-reversal symmetry due to the anti-unitarity of $\Theta$.
The resulting conditions on the parameters of the Hamiltonian following from Eq.~\eqref{eq:ch_symm} are again given in Table~\ref{tab:symm}.

\subsection{Particle-hole symmetry}
\label{Sec_3PH}

\begin{table*}
\caption{Conditions on pairwise couplings and on-site potentials for time-reversal, chiral, and particle-hole symmetry, which follow from Eqs.~\eqref{eq:tr_symm},~\eqref{eq:ch_symm},~\eqref{eq:ph_symm} and the corresponding choice of the $S_8$ or $S_1$ operator. The top and bottom row of each segment of the table must be identical. 
The sign in the first two relations for time-reversal symmetry and particle-hole symmetry with $\Pi=K S_8$ coincides with the sign of the relations $\Theta^2=\pm 1$,
$\Pi^2=\pm 1$.
Note that we allow for $J_{s \circ s'}(t) \in \mathbb C$ but, due to Hermiticity of the Hamiltonian, have $\Delta_s (t)\in \mathbb R$.
}
\begin{center}
\begin{ruledtabular}
\begin{tabular}{cccccccc}
\multicolumn{8}{c}{time-reversal symmetry} \\[2pt]
$J_{A\rightarrow B}(T-t)$ & $J_{C\rightarrow D}(T-t)$ & $J_{B\nearrow D}(T-t)$ & $J_{B\searrow D}(T-t)$ & $J_{D\nearrow B}(T-t)$ & $J_{D\searrow B}(T-t)$ & $\Delta_B(T-t)$ & $\Delta_D(T-t)$
\\
 $\pm J_{A\rightarrow B}(t)$   & $\pm J_{C\rightarrow D}(t)$ & $J_{A\nearrow C}^*(t)$ & $J_{A\searrow C}^*(t)$ & $J_{C\nearrow A}^*(t)$ & $J_{C\searrow A}^*(t)$ & $\Delta_A(t)$ & $\Delta_C(t)$
 \\[5pt] \hline
 \multicolumn{8}{c}{chiral symmetry} \\[2pt]
$J_{A\rightarrow B}(T-t)$ & $J_{C\rightarrow D}(T-t)$ & $J_{B\nearrow D}(T-t)$ & $J_{B\searrow D}(T-t)$ & $J_{D\nearrow B}(T-t)$ & $J_{D\searrow B}(T-t)$ & $\Delta_B(T-t)$ & $\Delta_D(T-t)$
\\
 $J_{A\rightarrow B}^*(t)$   & $J_{C\rightarrow D}^*(t)$ & $-J_{A\nearrow C}(t)$ & $J_{A\searrow C}(t)$ & $-J_{C\nearrow A}(t)$ & $J_{C\searrow A}(t)$ & $-\Delta_A(t)$ & $-\Delta_C(t)$
 \\[5pt] \hline
 \multicolumn{8}{c}{particle-hole symmetry with $\Pi=K S_1$} \\[2pt]
$J_{A\rightarrow B}(t)$ & $J_{C\rightarrow D}(t)$ & $J_{s\nearrow s'}(t)$ & $J_{s\searrow s'}(t)$ & $\Delta_A(t)$ & $\Delta_B(t)$ & $\Delta_C(t)$ & $\Delta_D(t)$
\\
$J_{A\rightarrow B}^*(t)$ & $J_{C\rightarrow D}^*(t)$ & $J_{s\nearrow s'}^*(t)$ & $J_{s\searrow s'}^*(t)$ & $0$ & $0$ & $0$ & $0$
 \\[5pt] \hline
 \multicolumn{8}{c}{particle-hole symmetry with $\Pi=K S_8$} \\[2pt]
$J_{A\rightarrow B}(t)$ & $J_{C\rightarrow D}(t)$ & $J_{B\nearrow D}(t)$ & $J_{B\searrow D}(t)$ & $J_{D\nearrow B}(t)$ & $J_{D\searrow B}(t)$ & $\Delta_B(t)$ & $\Delta_D(t)$
\\
 $\mp J_{A\rightarrow B}(t)$   & $ \mp J_{C\rightarrow D}(t)$ & $-J_{A\nearrow C}^*(t)$ & $-J_{A\searrow C}^*(t)$ & $-J_{C\nearrow A}^*(t)$ & $-J_{C\searrow A}^*(t)$ & $-\Delta_A(t)$ & $-\Delta_C(t)$
\end{tabular}
\end{ruledtabular}
\end{center}
\label{tab:symm}
\end{table*}

The symmetry relation for particle-hole symmetry is
\begin{equation}\label{eq:ph_symm}
 H_\mathrm{ph}(t) = - \Pi H_\mathrm{ph}(t) \Pi^{-1} \;,
\end{equation}
with an anti-unitary operator $\Pi$ for which $\Pi^2=\pm 1$.
Note that the same time argument $t$ appears on both sides of the relation.

For $\Pi^2=1$, use of the operator $S_8$
(with $\sigma=\tau=\sigma_x$)
forbids the appearance of horizontal pairwise couplings $A \to B$ and $C \to D$ in the driving protocol
according to the constraints 
listed in Tab.~\ref{tab:symm}.
Then, the lattice decouples into two independent (``red'' and ``blue'') sublattices. 
To avoid this situation, we use the operator $S_1$ for particle-hole symmetry with $\Pi^2=1$.

We now choose $\sigma = - \tau = \sigma_z$ with the Pauli matrix $\sigma_z$, such that
\begin{equation}
 S_1 =  n_A - n_B -  n_C + n_D \;,
\end{equation}
or $\sigma_{AA} = - \sigma_{BB} = - \tau_{CC}  =  \tau_{DD} = 1$.
The resulting conditions on the parameters of the Hamiltonian following from Eq.~\eqref{eq:ph_symm}, especially $\Delta_s(t) = 0$ for all on-site potentials, are given in Table~\ref{tab:symm}.

For $\Pi^2 = -1$, we have to use the operator $S_8$ according to the analysis in Sec.~\ref{sec:symm}.
We can choose $\sigma = \tau = \sigma_y$ as for fermionic time-reversal symmetry.
Now, however, the symmetry relation~\eqref{eq:ph_symm} contains the same time argument. The resulting conditions on the parameters of the Hamiltonian are again given in Table~\ref{tab:symm}.

\begin{figure*}
\hspace*{\fill}
\includegraphics[scale=0.8]{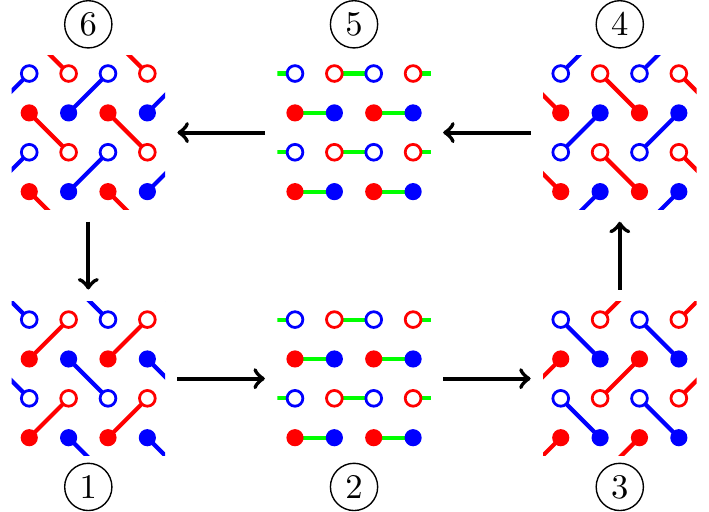}
\hspace*{\fill}
\includegraphics[scale=0.8]{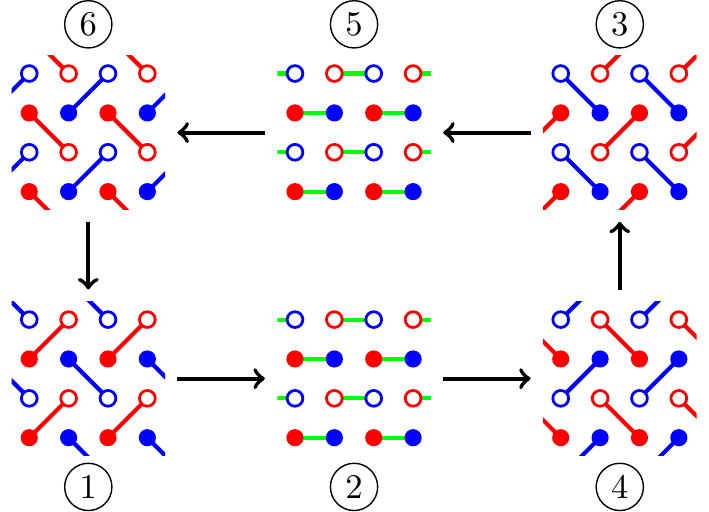}
\hspace*{\fill}
\caption{Two variants of the driving protocol \textsf{A},
which consist of a cyclic six-step sequence of the first five coupling patterns (a)---(e) from Fig.~\ref{fig:Lattice}.
Left panel: In this variant, the protocol consists of the sequence (b)$\to$(a)$\to$(c)$\to$(d)$\to$(a)$\to$(e).
Right panel: 
In this variant, the protocol consists of the sequence (b)$\to$(a)$\to$(d)$\to$(c)$\to$(a)$\to$(e).
As shown in the text, both variants are equivalent.
}
\label{Fig:six_step_protocol}
\end{figure*}

\begin{figure*}
\hspace*{\fill}
\includegraphics[scale=0.8]{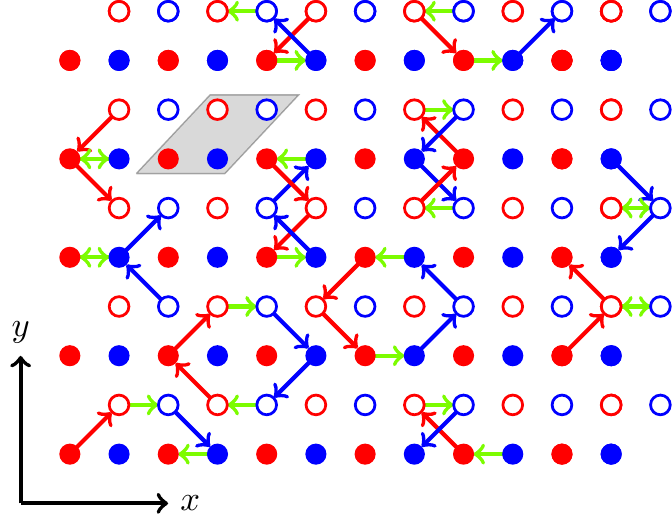}
\hspace*{\fill}
\includegraphics[scale=0.8]{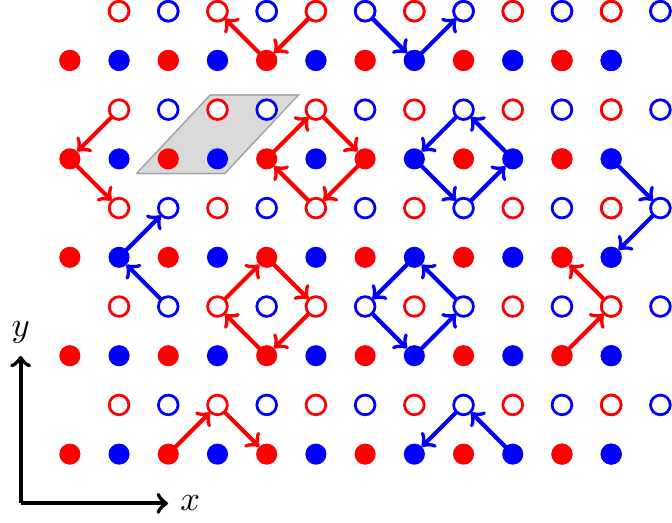}
\hspace*{\fill}
\caption{Patterns of motion during one cycle of driving protocol \textsf{A} at perfect coupling,
on a finite lattice of $6 \times 4$ unit cells (one unit cell is shown as a gray rhomboid).
The lattice comprises only entire unit cells,
such that the boundaries are compatible with the symmetry operators $S_1$, $S_8$.
The left and right panel correspond to the two variants of the protocol in Fig.~\ref{Fig:six_step_protocol}.
For the ``left'' variant the coupling in the horizontal steps 2,~5 is equal to $\pm J_p$,
for the ``right'' variant it is equal to zero.
}
\label{Fig:sketch_perfect}
\end{figure*}

\section{Universal driving protocol: Principal considerations}
\label{sec:principal}

The driving protocols considered here consist of $n$ consecutive steps
during which the Hamiltonian is constant.
Since we can always multiply the Hamiltonian in one step  by a number proportional to the step length,
we can assume that all steps have equal length $\delta t = T/n$, where $T$ is the period of the driving protocol.

Due to the constraints imposed on the Hamiltonian in Sec.~\ref{Sec_2},
each step of the driving protocol is given by one pattern of compatible pairwise couplings, several of which are shown in Fig.~\ref{fig:Lattice}.
While there remains some ambiguity in the construction of the protocol, the selection of the coupling patterns, and their arrangement into the $n$-step sequence,
has to be carried out according to the symmetry analysis from Sec.~\ref{sec:symm}.
In particular, only coupling patterns that are compatible with the symmetry operator $S_8$ can be chosen in the construction.

The principal distinction between the two protocols that will be introduced in Secs.~\ref{Sec_4},~\ref{sec:PH} arises from the time argument in the symmetry relations~\eqref{eq:tr_symm},~\eqref{eq:ph_symm}.
For time-reversal symmetry, where different time arguments $t$ and $T-t$ appear on either sides of the symmetry relation~\eqref{eq:tr_symm}, in principal any pattern ``(p)'' compatible with $S_8$ can be used in the protocol if its counterpart ``$S_8 \, \mathrm{(p)} \, S_8^{-1}$'' appears at $T-t$.
Exploration of the different combinations quickly shows that only
the four patterns (b)--(e) with perpendicular diagonal couplings give rise to a non-trivial driving protocol with a small number of steps.
In fact, it is not surprising that perpendicular couplings should be used since the protocol has to support counter-propagating boundary states for time-reversal symmetry (see also App.~\ref{app:Parallel}).
Therefore, the driving protocol for time-reversal symmetry (``protocol \textsf{A}'') will be constructed out of the four patterns (b)--(e) in Fig.~\ref{fig:Lattice} with perpendicular diagonal couplings, in combination with the horizontal pattern (a).

For particle-hole symmetry, where the same time argument $t$ appears on both sides of the symmetry relation~\eqref{eq:ph_symm}, only patterns that are mapped onto themselves by $S_8$ can be used.
Therefore, the driving protocol for particle-hole symmetry (``protocol \textsf{B}'') will be constructed out of patterns with parallel diagonal couplings (pattern (f) in Fig.~\ref{fig:Lattice}, or patterns (f1)--(f4) in  Fig.~\ref{fig:para_diag} in the appendix), in combination with the horizontal pattern~(a).

\section{Universal driving protocol \textsf{A}: Time-reversal symmetry}
\label{Sec_4}

\subsection{Construction of the protocol}
\label{sec:mini:a}

According to the previous section, we construct the driving protocol \textsf{A} for time-reversal symmetry
out of the four perpendicular diagonal coupling patterns (b)--(e) from Fig.~\ref{fig:Lattice}.
How these patterns should be arranged into a sequence can now be deduced from the mappings induced by the operator $S_8$.
If we start the sequence with, say, pattern (b), the sequence has to end with pattern (e) since $S_8$ swaps (b)$\leftrightarrow$(e). If the second step in the sequence is pattern (c), the penultimate step in the sequence must be pattern (d) since $S_8$ swaps (c)$\leftrightarrow$(d).
Therefore, only two different four-step sequences qualify for driving protocol \textsf{A}:
(b)$\to$(c)$\to$(d)$\to$(e) and
(b)$\to$(d)$\to$(c)$\to$(e).
Starting with different patterns results in equivalent sequences.

In these two four-step sequences,
the ``red'' and ``blue'' sublattice of the square lattice remain decoupled, as can be deduced from Fig.~\ref{fig:Lattice}. Therefore, a horizontal (or, equivalently, vertical) coupling pattern has to be added to the sequence. The only pattern of this type compatible with $S_8$ is pattern (a) in Fig.~\ref{fig:Lattice}.
In order to allow for time-reversal or chiral symmetry, pattern (a) has to appear in symmetric position in the sequence:
(i) as the central step 3 of a five-step sequence,
(ii) as steps 1,~6 or (iii) steps 2,~4 of a six-step sequence.
Taking into account that according to Table~\ref{tab:transform} fermionic time-reversal symmetry changes the sign of the parameters $J_{A \rightarrow B}$, $J_{C \rightarrow D}$ of pattern (a),
only the last possibility (iii) results in a non-trivial addition to the sequence.

To summarize, we have the two variants of driving protocol \textsf{A} shown in Fig.~\ref{Fig:six_step_protocol}.
The protocol is constructed out of the first five coupling patterns in Fig.~\ref{fig:Lattice},
and according to our construction will be able to support either time-reversal, chiral, or particle-hole symmetry.
In each step, the coupling patterns can be combined with arbitrary on-site potentials $\Delta_s(t)$ without changing the structure of the driving protocol or violating the constraints on the Hamiltonian.
This gives a total of $6 \times (2 + 4) = 36$ parameters,
which are further restricted by the conditions in Table~\ref{tab:symm} if the respective symmetry is enforced.

\subsection{Perfect coupling}

``Perfect coupling'' denotes the situation where all on-site potentials $\Delta_s \equiv 0$,
and the pairwise couplings in a given step are either both $J_{s \circ s'} \equiv 0$ or 
$|J_{s \circ s'}| \equiv J_p$, with $J_p = \pi/ (2 \, \delta t)$ (here, for six steps, $J_p= 3 \pi /T$).
The sign of the $J_{s \circ s'}$ parameters must be chosen according to Table~\ref{tab:symm} for the respective symmetry.

At perfect coupling, pairwise coupling fully transfers the amplitude on one lattice site to an adjacent lattice site. The driving protocol reduces to a sequence of jumps that follow the geometric shapes of the coupling patterns.

The resulting patterns of motion for the two variants of driving protocol \textsf{A} are shown in Fig.~\ref{Fig:sketch_perfect}.
The difference between the two variants is only the coupling in the horizontal steps 2~and~5,
which is equal to $\pm J_p$ for the ``left'' variant and equal to zero for the ``right'' variant.
A particle in the bulk moves in a closed loop,
while a particle at the boundary is transported by two sites in one cycle. The direction of motion depends on the starting site (``red'' or ``blue''). This pattern of motion gives rise to a non-trivial topological phase, and to a symmetry-protected pair of boundary states with opposite chirality. 

Note that when we introduce boundaries, either here or for Figs.~\ref{fig:perfect_boundary}--\ref{Fig:edge_state_prop} below, we always choose boundaries that do not separate sites within one unit cell,
and thus are compatible with the symmetry operators $S_1$ and $S_8$.
As   in Fig.~\ref{Fig:sketch_perfect}, boundaries along the $x$-direction ($y$-direction) are parallel to the translation vector $\vec a_x$ ($\vec a_y$).

For perfect coupling,
scattering between boundary states with opposite chirality is strictly forbidden by the construction of the protocol, rather than by a topological constraint.
In particular, a state starting on a ``red'' (``blue'') site always ends up on a ``red'' (``blue'') site after a full cycle.
Fully developed symmetry-protected phases require general parameters in the driving protocol, and will be studied in the next section.

\subsection{Equivalence of driving protocols}
\label{sec:EquivalentProtocols}

The patterns of motion in Fig.~\ref{Fig:sketch_perfect} suggest that the two variants of the driving protocol \textsf{A} are in fact equivalent.
As we show now, the equivalence holds not only at perfect coupling but in general.

The Floquet propagator $U(T)$, over one period of the driving protocol, is a simple product
\begin{equation}
 U(T) = U_6 U_5 U_4 U_3 U_2 U_1
\end{equation}
of the Floquet propagators $U_k = \exp[-\ii \delta t H_k]$ for each of the steps $k=1,..,6$,
with constant Hamiltonian $H(t) \equiv H_k$ for $(k-1) \delta t \le t \le k \delta t$ in step $k$.

Now let $S = \hat t_{A \rightarrow B} + \hat t_{A \rightarrow B}^\dagger +
\hat t_{C \rightarrow D} + \hat t_{C \rightarrow D}^\dagger $
be the unitary operator that swaps the ``red'' and ``blue'' sublattice (we have $S =  S^+$ and $S^2 = 1$).
In fact, $S$ is a special case of the symmetry operator $S_8$,
and $S = - \ii{} U_2 = - \ii{} U_5$ at perfect coupling $J_{A \rightarrow B} = J_{C \rightarrow D} = J_p$.

Inserting $S$ into the Floquet propagator, we have the alternative expression
\begin{equation}
 U(T) = U_6 \, (U_5 S^\dagger) \, (S U_4 S^\dagger) \, (S U_3 S^\dagger) \, (S U_2) \, U_1 \;.
\end{equation}
Since $S$ swaps the ``red'' and ``blue'' sublattice, it effectively exchanges steps 3 and 4.
On the other hand, the product $S U_2$ can be combined into a horizontal coupling step 2 with modified parameters, as in
\begin{equation}
 S \, U_2 \begin{bmatrix} J_{A \rightarrow B} \\  J_{C \rightarrow D} \end{bmatrix} = - \ii\, U_2
 \begin{bmatrix} J_{A \rightarrow B} - J_p \\  J_{C \rightarrow D} - J_p \end{bmatrix} 
\;,
\end{equation}
where we include the coupling parameters explicitly.

Therefore, the Floquet propagator $U(T)$, over one driving period of the protocol,
is identical (up to a sign $(- \ii)^2 = -1$) for both variants if the parameters 
of the horizontal coupling steps 2,~5 are modified by $\pm J_p$ according to the above transformation.
Especially at perfect coupling, the parameters are either $J_p$ (``left'' variant) or zero (``right'' variant), as in Fig.~\ref{Fig:sketch_perfect}.

Note that the ``right'' variant in Fig.~\ref{Fig:sketch_perfect} has a close connection to the driving protocol from Ref.~\cite{PhysRevX.3.031005}, which realizes Floquet topological insulators without additional symmetries.
Essentially, two copies of this protocol have to be combined to obtain our driving protocol with symmetries.
The details of the combination, as well as the conditions on the protocol parameters,
follow from the symmetry analysis provided here.

\subsection{Equivalence of coupling steps}
\label{sec:Circumvent}

Similar to the entire driving protocol,
also the individual steps can be written in different equivalent ways.
To see how, assume that the Hamiltonian in one step of duration $\delta t$ is of the form $H_{\mathrm{step}}=J (\hat t_{s \circ s'} + \hat t_{s \circ s'}^\dagger)+\Delta (\hat n_s -  \hat n_{s'})$,
with $J, \Delta \in \mathbb R$. The propagator for this step evaluates to
\begin{equation}
\label{U_step}
\begin{split}
 U_\mathrm{step}(J, \Delta) & = \exp [ - \ii\, \delta t  \, H_{\mathrm{step}} ] \\
  & = \cos(\delta t  \, \xi  ) \, \mathbbm 1 \,- \, \ii \frac{\sin(\delta t \, \xi  )}{\xi} \, H_{\mathrm{step}} \;,
  \end{split} 
\end{equation}
with $\xi =(J^2+\Delta^2)^{1/2}$.
Essentially, this propagator is an $\mathrm{SU}(2)$ rotation.

The right hand side of Eq.~\eqref{U_step} is periodic in the quantity $\xi$. Therefore,
\begin{equation}\label{U_step_mod}
 U_\mathrm{step}(J, \Delta) = (- 1)^m \, U_\mathrm{step}(\alpha_m J , \alpha_m \Delta)
\end{equation}
for every $\alpha_m = 1 + (2 m J_p) / (J^2 + \Delta^2)^{1/2}$ with $m \in \mathbb Z$.

This relation becomes especially clear for $\Delta = 0$, where
$U_\mathrm{step}(J, 0) = (-1)^m \, U_\mathrm{step}(J + 2 m J_p, 0)$.
In particular for perfect coupling $|J| = J_p = \pi/(2 \, \delta t)$, 
where $U(\pm J_p , 0) = \mp \ii (\hat t_{s \circ s'} + \hat t_{s \circ s'}^\dagger)$,
negative and positive couplings $J = \pm J_p$ are equivalent. 

The equivalence of coupling steps with different parameters has important consequences,
both conceptually (see Sec.~\ref{sec:switch}) as well as practically for a photonic lattice implementation.
Implementation of negative couplings between waveguides is a challenging procedure~\cite{PhysRevLett.116.213901}, but depending on the symmetry negative couplings cannot be avoided in the driving protocol (cf. Table~\ref{tab:symm}).
Fortunately, any negative coupling $J < 0$ can be replaced by an equivalent positive coupling $\alpha_m J$ from Eq.~\eqref{U_step_mod}.
This argument shows that negative couplings are not a principal obstacle against a photonic lattice implementation of the universal driving protocol.

\pagebreak

\section{Symmetry-protected Floquet topological phases}
\label{Sec_5}

\begin{figure}[b]
\hspace*{\fill}
\includegraphics[width=0.45\linewidth]{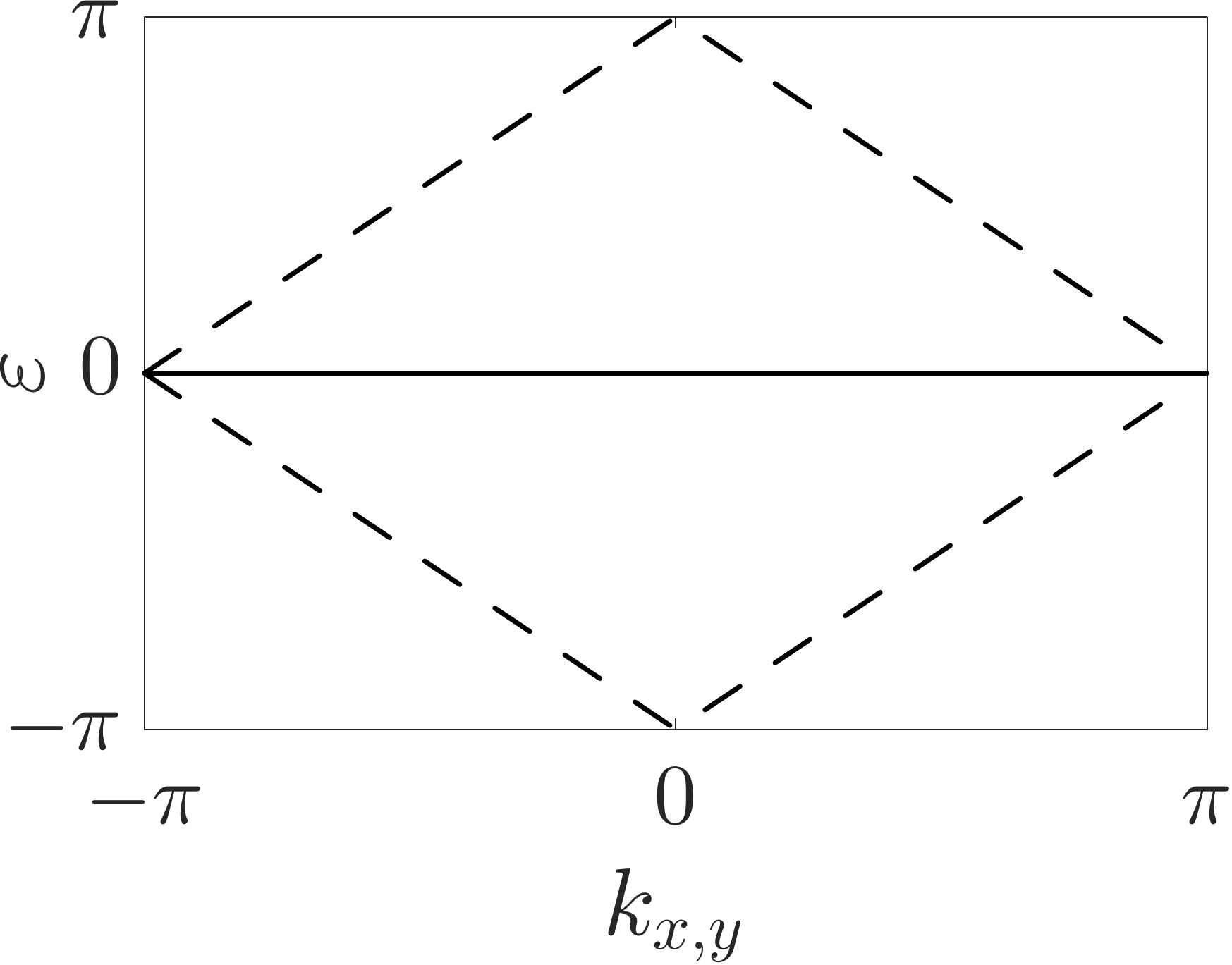}
\hspace*{\fill}
\includegraphics[width=0.45\linewidth]{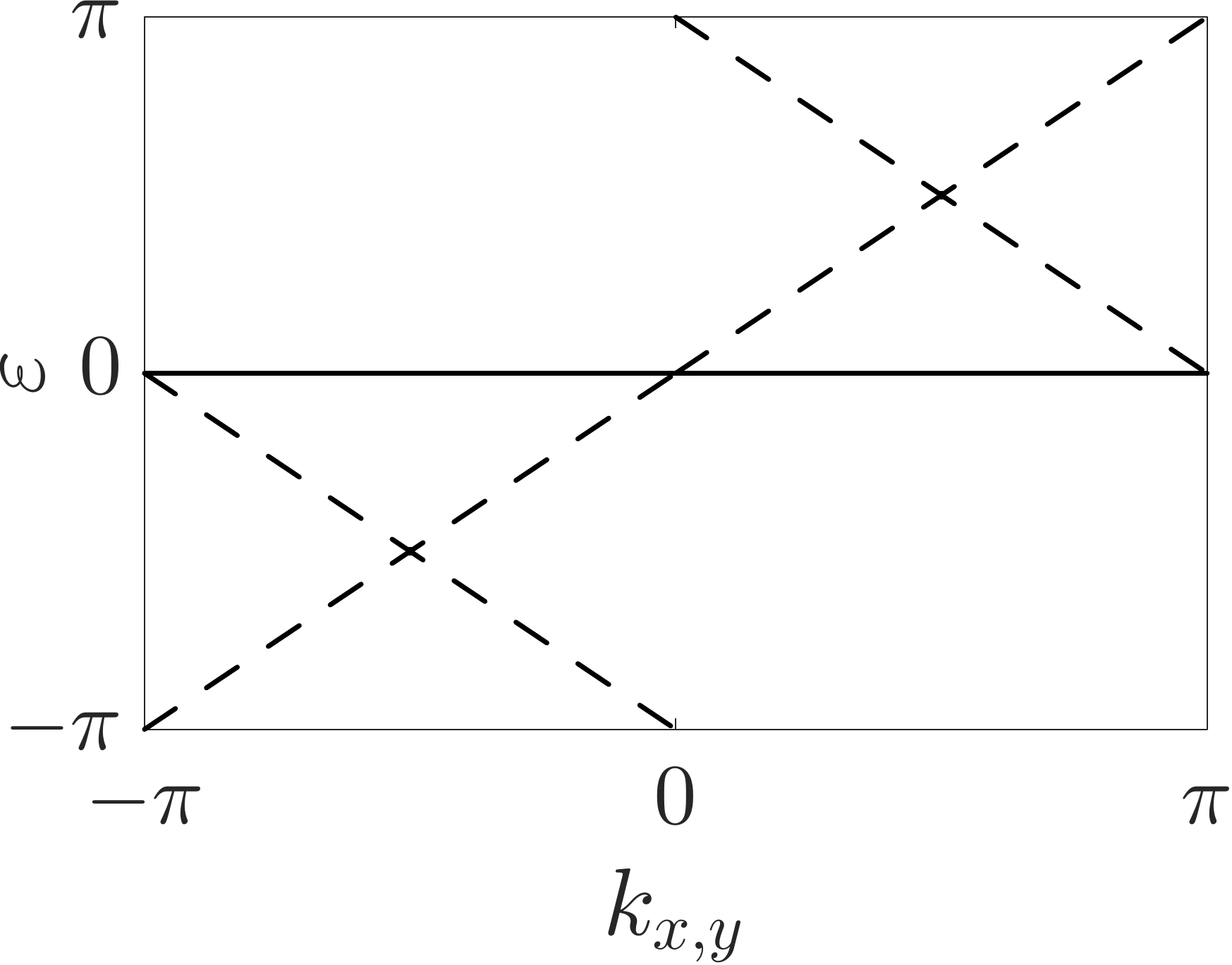}
\hspace*{\fill}
\caption{Dispersion of bulk (solid) and boundary (dashed) states for perfect coupling with fermionic time-reversal (left panel) or chiral (right panel) symmetry.
Parameter values can be deduced from the corresponding columns in Table~\ref{Tab:non_perfect},
setting $J = J_p$ and $\Delta = 0$.
Here and in Figs.~\ref{Fig:non_perfect1}--\ref{fig:ch_ph} we show the states on one boundary of a semi-infinite ribbon, and do not include the states on the opposite boundary.
}
\label{fig:perfect_boundary}
\end{figure}

\begin{table}
\caption{Parameter sets for driving protocol \textsf{A} with time-reversal (TRS), chiral (CS), or particle-hole symmetry (PHS). 
TRS and CS have the two free parameters $J$, $\Delta$.
PHS with $\Pi^2 = 1$ has two free parameters $J$, $J'$.
Perfect coupling corresponds to $\Delta = 0$ and $J = J' = J_p$,
where $J_p = 3 \pi/T$ for  a six-step protocol.
Unspecified parameters are zero,
and the sign in step 5 of the TRS column is $+$ for bosonic and $-$ for fermionic time-reversal symmetry.
In Figs.~\ref{Fig:non_perfect1}--\ref{Fig:edge_state_prop}, we use the values of $\Delta, J, J'$ specified under ``this work''.
}
\begin{center}
\begin{ruledtabular}
\begin{tabular}{llll}
 & $\mathrm{TRS}$ & $\mathrm{CS}$ & $\mathrm{PHS} \; \Pi^2=1$
\\ \hline
 step $1$ &  $J_{A\nearrow C}=J_{\mathrm{p}}$ & $J_{A\nearrow C}=J_{\mathrm{p}}$ & $J_{A\nearrow C}=J_{\mathrm{p}}$ \\
&  $J_{B\searrow D}=J_{\mathrm{p}}$ & $J_{B\searrow D}=J_{\mathrm{p}}$ & $J_{B\searrow D}=J_{\mathrm{p}}$ \\
 & $\Delta_{A}=\Delta_{B}= \phantom{-} \Delta$ & $\Delta_{B}=- \Delta$ &\\
 & $\Delta_{C}=\Delta_{D}= - \Delta$ & $\Delta_{D}= \phantom{-} \Delta$ &\\[3pt] \cline{2-4}

step $2$ & $J_{A\rightarrow B}=J$ & $J_{A\rightarrow B}=J$ & $J_{A\rightarrow B}=J$ \\
& $J_{C\rightarrow D}=J$ & $J_{C\rightarrow D}=J$ & $J_{C\rightarrow D}=J$ \\[3pt] \cline{2-4}

 step $3$ &  $J_{C\nearrow A}=J_{\mathrm{p}}$ & $J_{C\nearrow A}=J_{\mathrm{p}}$ & $J_{C\nearrow A}=J_{\mathrm{p}}$ \\
&  $J_{D \searrow B}=J_{\mathrm{p}}$ & $J_{D \searrow B}=J_{\mathrm{p}}$ & $J_{D \searrow B}=J_{\mathrm{p}}$ \\
 & $\Delta_{A}=\Delta_B= \phantom{-} \Delta$ & &\\
 & $\Delta_{C}=\Delta_D=- \Delta$ & &\\[3pt] \cline{2-4}
 step $4$ &  $J_{C\searrow A}=J_{\mathrm{p}}$ & $J_{C\searrow A}=\phantom{-}  J_{\mathrm{p}}$ & $J_{C\searrow A}= \phantom{-} J_{\mathrm{p}}$ \\
&  $J_{D\nearrow B}=J_{\mathrm{p}}$ & $J_{D\nearrow B}=-J_{\mathrm{p}}$ & $J_{D\nearrow B}=-J_{\mathrm{p}}$ \\
 & $\Delta_{A}=\Delta_B=\phantom{-}  \Delta$ & &\\
 & $\Delta_{C}=\Delta_D=- \Delta$ & &\\[3pt] \cline{2-4}

step $5$ & $J_{A\rightarrow B}=\pm J$ & $J_{A\rightarrow B}=J$ & $J_{A\rightarrow B}=J'$ \\
& $J_{C\rightarrow D}=\pm J$ & $J_{C\rightarrow D}=J$ & $J_{C\rightarrow D}=J'$ \\[3pt] \cline{2-4}

 step $6$ &  $J_{A\searrow C}=J_{\mathrm{p}}$ & $J_{A\searrow C}=\phantom{-}  J_{\mathrm{p}}$ & $J_{A\searrow C}=\phantom{-}  J_{\mathrm{p}}$ \\
&  $J_{B\nearrow D}=J_{\mathrm{p}}$ & $J_{B\nearrow D}=-J_{\mathrm{p}}$ & $J_{B\nearrow D}=-J_{\mathrm{p}}$ \\
 & $\Delta_{A}=\Delta_{B}=\phantom{-}  \Delta$ & $\Delta_{A}=\phantom{-}  \Delta$ &\\
 & $\Delta_{C}=\Delta_{D}=- \Delta$ & $\Delta_{C}=- \Delta$ & 
 \\[3pt] \hline
this & $J = 2 \pi/T$ & $J = 2 \pi/T$ & $J = 2 \pi/T$ \\
work & $\Delta = 3/T$ & $\Delta = 9/T$ & $J' =\,\pi/T$
\end{tabular}
\end{ruledtabular}
\end{center}
\label{Tab:non_perfect}
\end{table}

In $2+1$ dimensions~\cite{PhysRevB.96.155118}, fermionic time-reversal symmetry ($\Theta^2=-1$) leads to a symmetry-protected $\mathbb Z_2$  topological phase with counterpropagating boundary states. Bosonic time-reversal symmetry ($\Theta^2=1$) does not lead to a non-trivial topological phase.
Particle-hole symmetry with $\Pi^2=1$ allows for generic Chern insulators without additional symmetry protection, while particle-hole symmetry with $\Pi^2=-1$ features a $2\mathbb Z$ topological phase with an even number of copropagating chiral boundary states.

The symmetry-protected $\mathbb Z_2$~phase with fermionic time-reversal symmetry is realized in the driving protocol~\textsf{A}. Since this constitutes the most interesting situation, we start with an extended discussion of topological phases and boundary states in this protocol.
The $2\mathbb Z$ phase with particle-hole symmetry will be discussed after the introduction of protocol~\textsf{B} in Sec.~\ref{sec:PH}.

At perfect coupling, the driving protocol \textsf{A} realizes a non-trivial topological phase with counterpropagating boundary states that follow the patterns of motion in Fig.~\ref{Fig:sketch_perfect}.
The bulk bands and boundary state dispersions are shown in Fig.~\ref{fig:perfect_boundary},
where we plot the Floquet quasienergies $\varepsilon$ as a function of momentum $k_x$ or $k_y$ along a boundary in $x$- or $y$-direction.
The quasienergies are computed from the eigenvalues $e^{-\ii \varepsilon}$ of the Floquet propagator after one driving period.
At perfect coupling, the bulk bands are flat at $\varepsilon = 0$ and a gap exists at $\varepsilon = \pi$.
The boundary states have linear dispersion,
which does not depend on the orientation of the boundary.
Due to symmetry, they occur in pairs of opposite chirality. 
Furthermore, with zero potential $\Delta_s \equiv 0$, time-reversal or chiral symmetry appears together with particle-hole symmetry.

To realize symmetry-protected Floquet topological phases away from perfect coupling,
we use the parameter values listed in Table~\ref{Tab:non_perfect}.
Of the $36$ parameters of the protocol, at most $28$ parameters are assigned non-zero values.
Steps 2,~5 do not involve on-site potentials, and all parameters are real.
It is straightforward to check that the three parameter sets fulfill either the conditions of time-reversal, chiral, or particle-hole symmetry in Table~\ref{tab:symm}.
Each set depends on two free parameters,
and includes the perfect coupling case in Fig.~\ref{fig:perfect_boundary}.
For the remainder of this section, we use the parameters listed under ``this work'', and the ``left'' variant of driving protocol \textsf{A} in Fig.~\ref{Fig:six_step_protocol}.

\subsection{Time-reversal symmetry}

\begin{figure*}
\hspace*{\fill}
\includegraphics[scale=0.8]{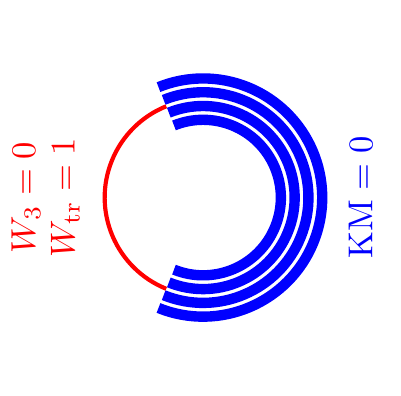}
\hspace*{\fill}
\includegraphics[scale=0.23]{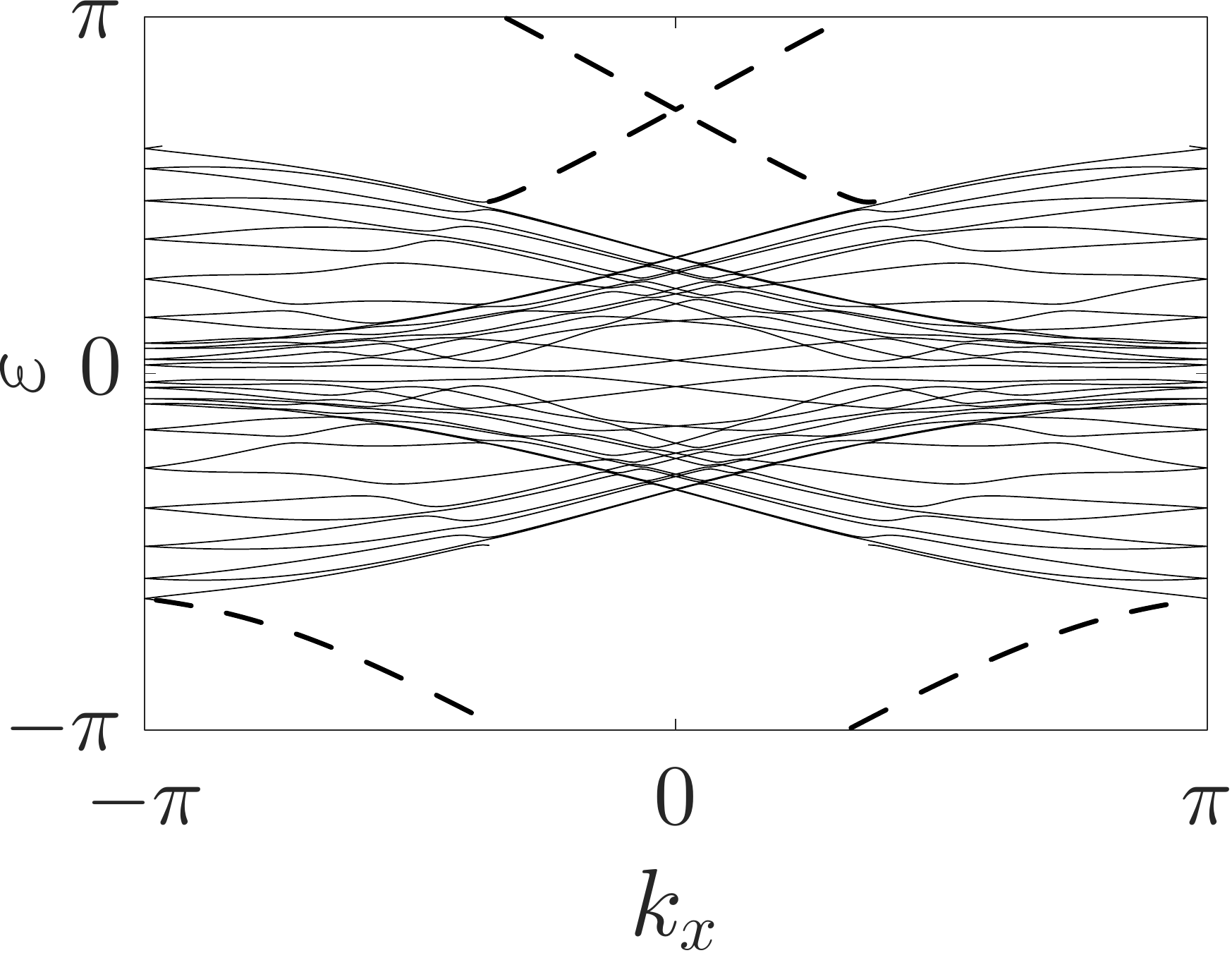}
\hspace*{\fill}
\includegraphics[scale=0.23]{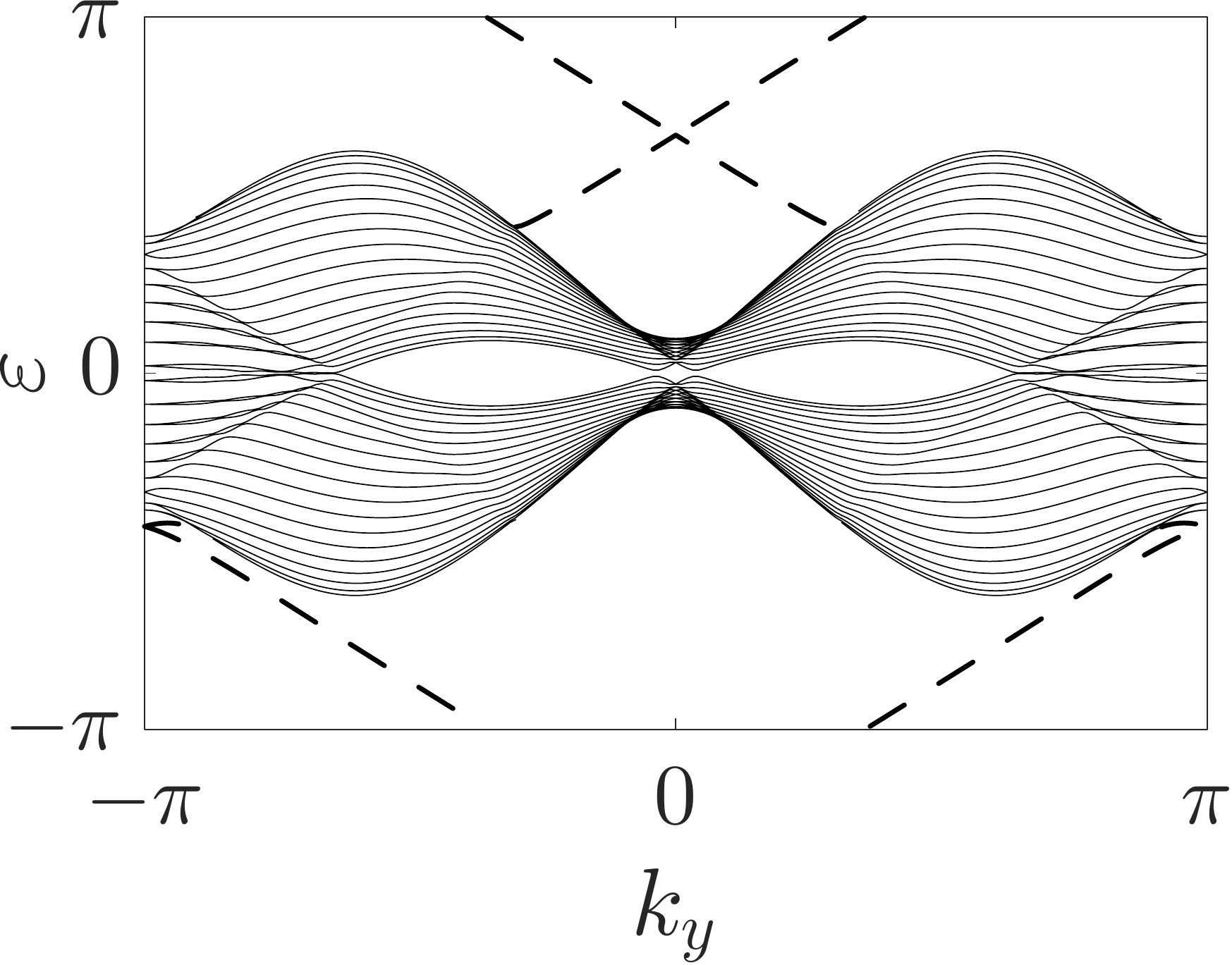}
\hspace*{\fill}
\\[1.73ex]
\hspace*{\fill}
\includegraphics[scale=0.8]{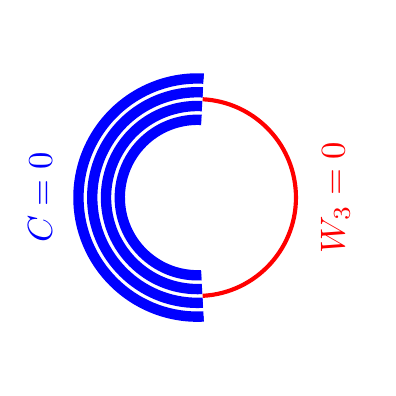}
\hspace*{\fill}
\includegraphics[scale=0.23]{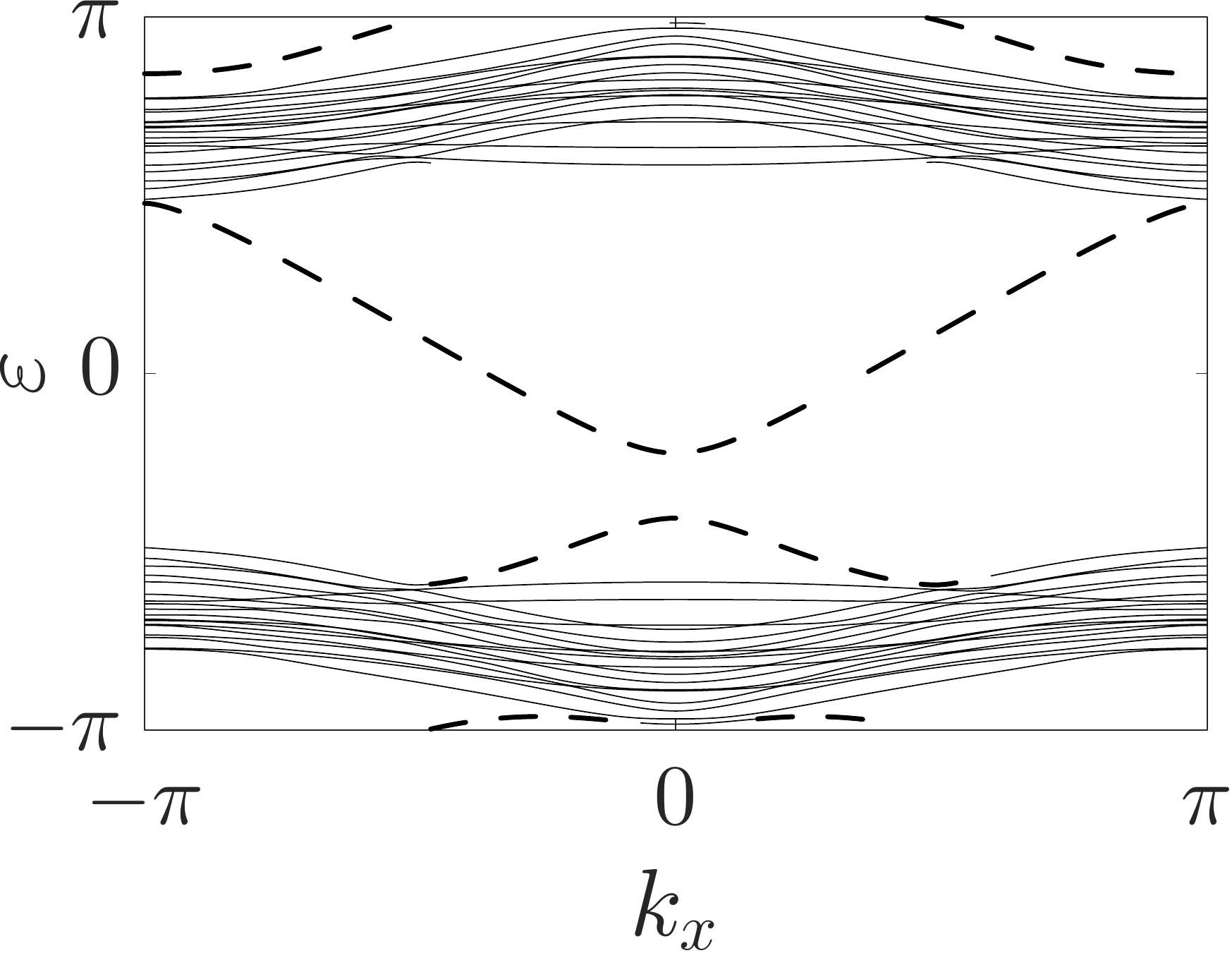}
\hspace*{\fill}
\includegraphics[scale=0.23]{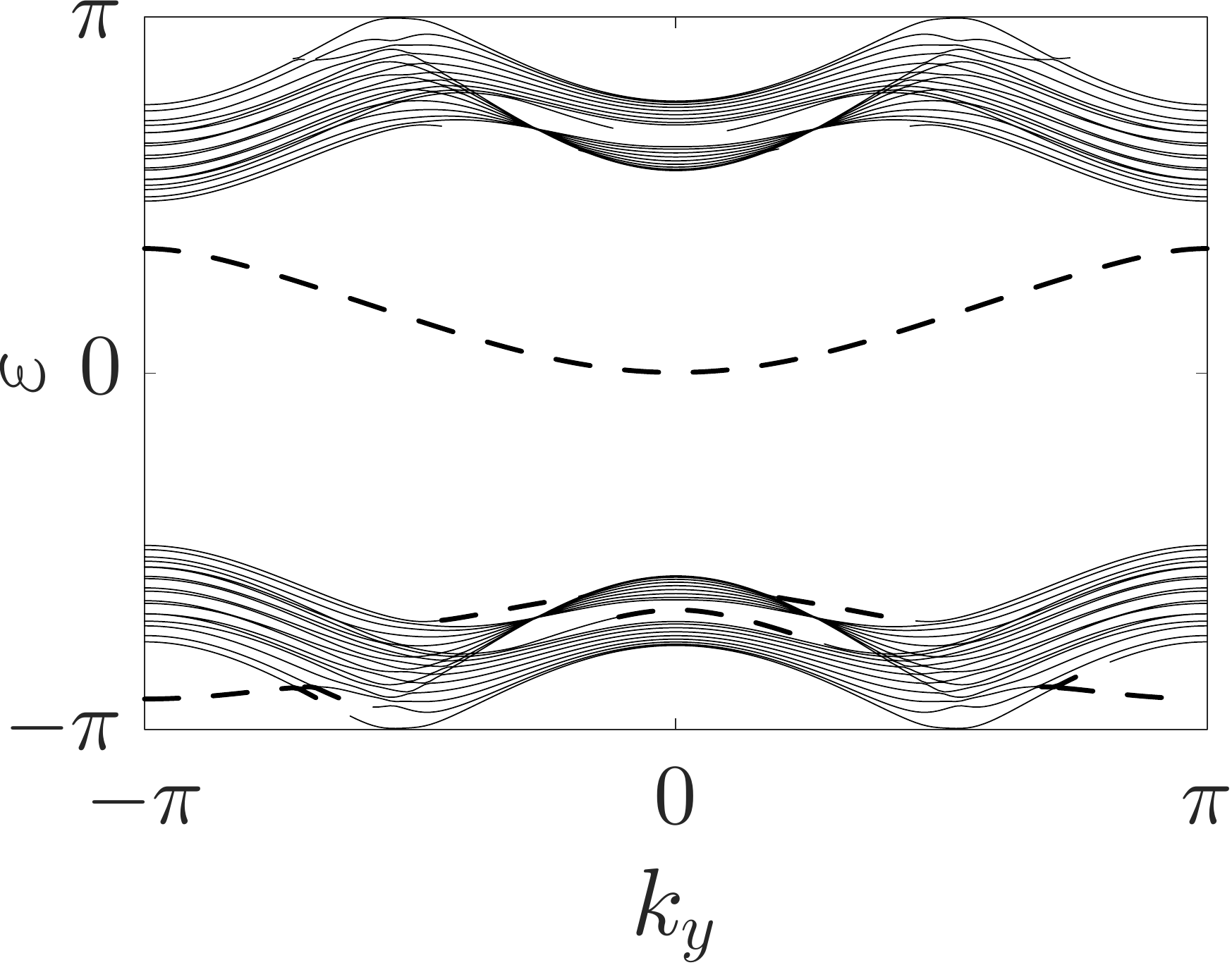}
\hspace*{\fill}
\caption{Floquet bands and boundary states for fermionic (top row) and bosonic (bottom row) time-reversal symmetry, using the parameters from Tab.~\ref{Tab:non_perfect}.
Left column: Blue arcs indicate the (four-fold degenerate) Floquet bands,
red arcs the gaps. Quasienergies $\varepsilon$ are plotted on the circle $\varepsilon \mapsto e^{-\ii \varepsilon}$. 
Included are the respective (Kane-Mele $\mathrm{KM}$ or Chern number $C$) invariants of the bands, and the 
$W_\mathrm{tr}$-invariant or the $W_3$-invariant associated with the gap.
Central and right column: Floquet bands (solid) and boundary state dispersion (dashed), as a function of momentum $k_x$ or $k_y$ for a boundary along the $x$- or $y$-direction. 
}
\label{Fig:non_perfect1}
\end{figure*}

In Fig.~\ref{Fig:non_perfect1} we show the Floquet bands and boundary states
for fermionic and bosonic time-reversal symmetry.
Both cases differ only by the sign of the parameters in step 5 of the driving protocol (cf. Table~\ref{Tab:non_perfect}), such that the gap is either at $\varepsilon=\pi$ (fermionic) or $\varepsilon=0$ (bosonic).

For fermionic time-reversal symmetry (top row in Fig.~\ref{Fig:non_perfect1}),
two boundary states with opposite chirality traverse the gap.
The crossing of the boundary states at the invariant momentum $k_{x,y} = 0$ is protected by Kramers degeneracy.
Since the two boundary states are mapped onto each other by the symmetry operator $S_8$,
they can be described as helical boundary states in the pseudo-spin interpretation of the driving protocol given in App.~\ref{app:Pseudo}.

Because of time-reversal symmetry, the boundary states have to appear in pairs of opposite chirality.
In this situation, the $W_3$-invariant~\cite{PhysRevX.3.031005}, which counts the net chirality of boundary states in a gap of a Floquet system, necessarily vanishes.
Therefore, the topological phase observed here is not protected against general deformations of the Floquet Hamiltonian, but only against deformations that preserve time-reversal symmetry.

To characterize this symmetry-protected topological phase we can compute the relevant $\mathbb Z_2$-valued bulk invariant~\cite{PhysRevLett.114.106806,1367-2630-17-12-125014,HAF18}.
In the present situation, we get a non-zero invariant ($W_\mathrm{tr} \ne 0$ in the notation of Ref.~\cite{HAF18},
computed with the algorithm from Ref.~\cite{HAF17}). This confirms that the driving protocol indeed supports a non-trivial time-reversal symmetric topological phase, with a pair of counterpropagating boundary states.

Additionally, we find that the Kane-Mele invariants~\cite{PhysRevLett.95.146802,PhysRevB.74.195312,doi:10.1143/JPSJ.76.053702} of the individual Floquet bands are zero. We recognize the signature of an anomalous Floquet topological phase~\cite{PhysRevX.3.031005,1367-2630-17-12-125014}, which exists although all Floquet bands are topologically trivial.

For bosonic time-reversal symmetry (bottom row in Fig.~\ref{Fig:non_perfect1}), the $W_3$-invariant still has to be zero.
Now, however, crossing of the boundary states is not protected by Kramers degeneracy.
The boundary states do not have to traverse the gap
and can be deformed continuously to merge with the Floquet bands, without breaking the symmetry.
Consequently, the system is 
topologically trivial.

\begin{figure}[b]
\hspace*{\fill}
\includegraphics[width=\linewidth]{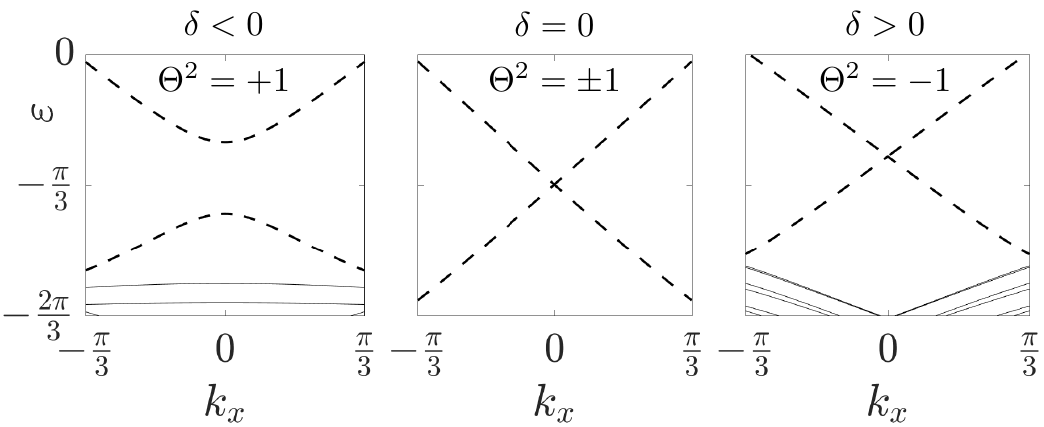}
\hspace*{\fill}
\caption{Switching between bosonic (left and central panel) and fermionic (central and right panel) time-reversal symmetry through continuous variation of the parameter $\delta$ (see text).
The panels show the boundary state dispersion. The protocol parameters for the negative and positive $\delta =\pm \pi/T$ used here agree with Fig.~\ref{Fig:non_perfect1}.
}
\label{fig:phase_trans}
\end{figure}

\begin{figure*}
\hspace*{\fill}
\includegraphics[scale=0.8]{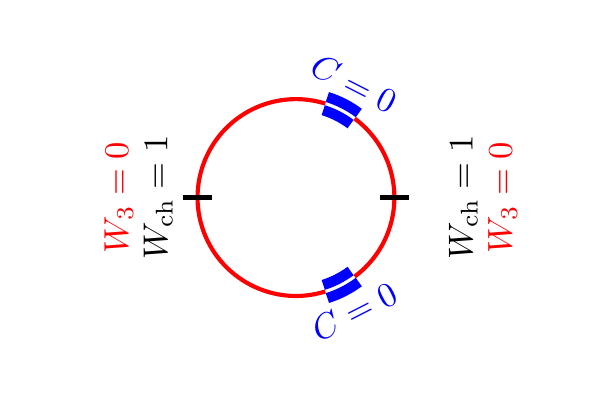}
\hspace*{\fill}
\includegraphics[scale=0.23]{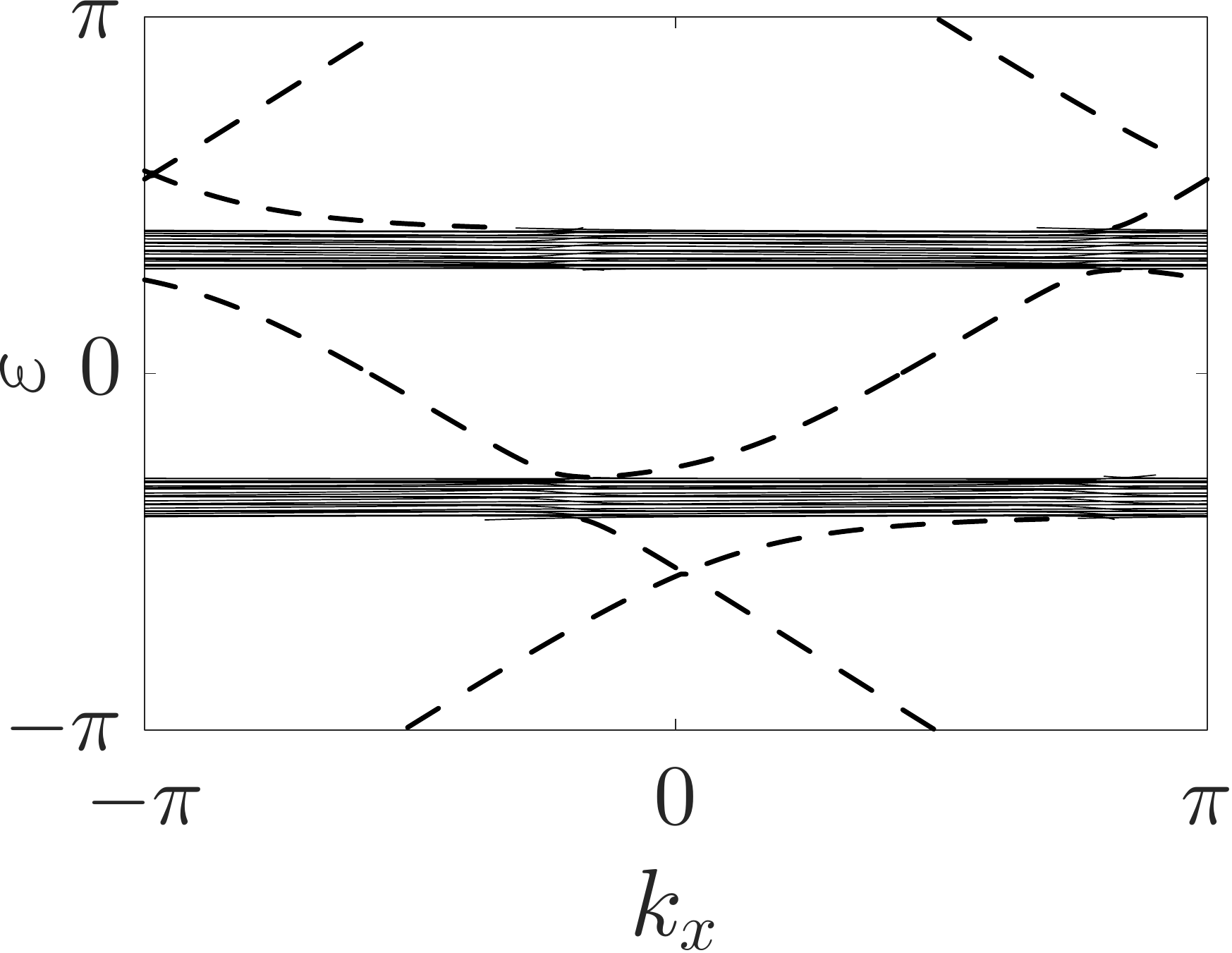}
\hspace*{\fill}
\includegraphics[scale=0.23]{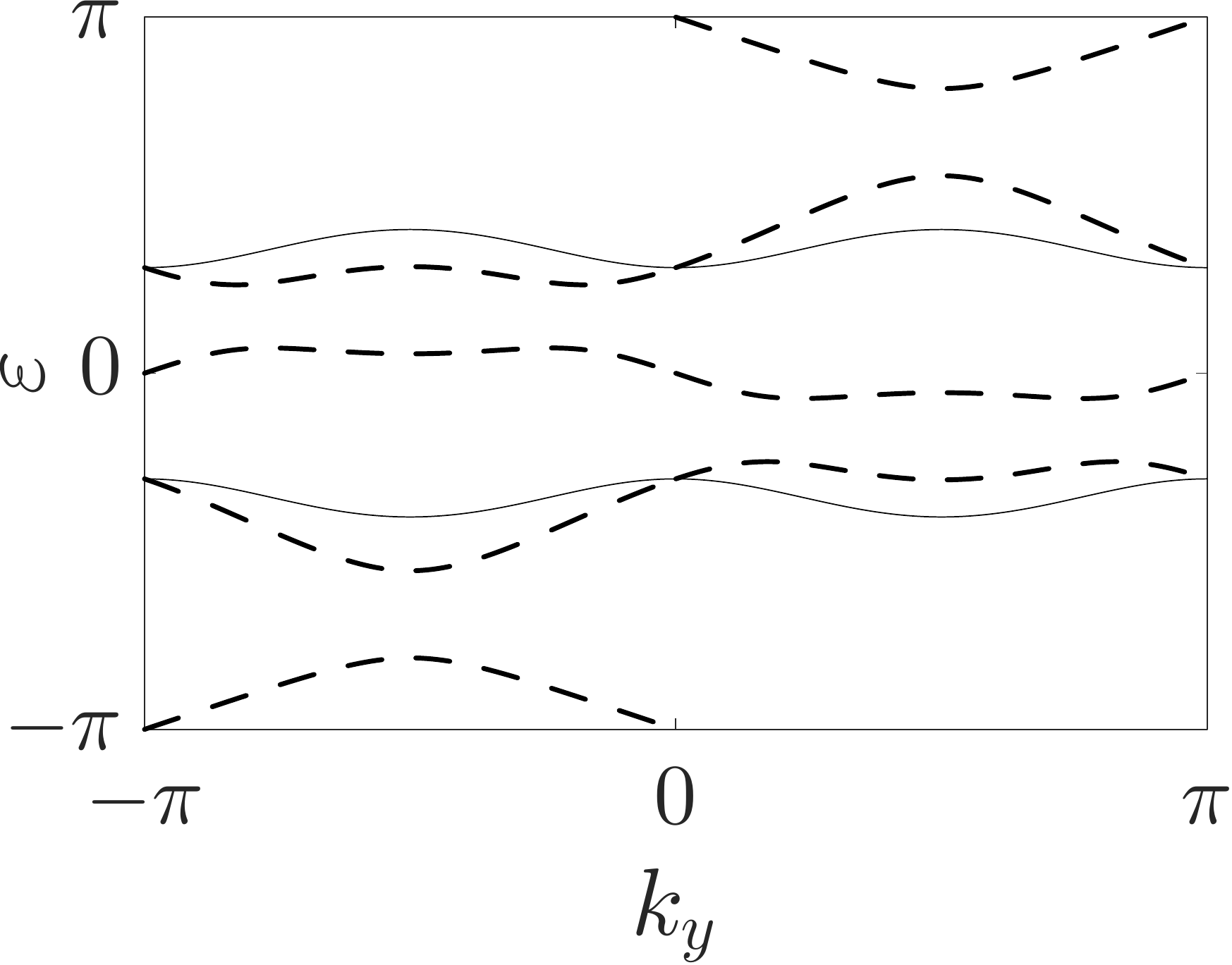}
\hspace*{\fill}
\\
\hspace*{\fill}
\includegraphics[scale=0.8]{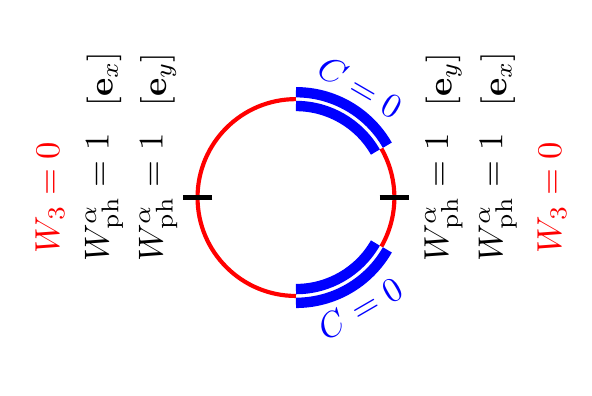}
\hspace*{\fill}
\includegraphics[scale=0.23]{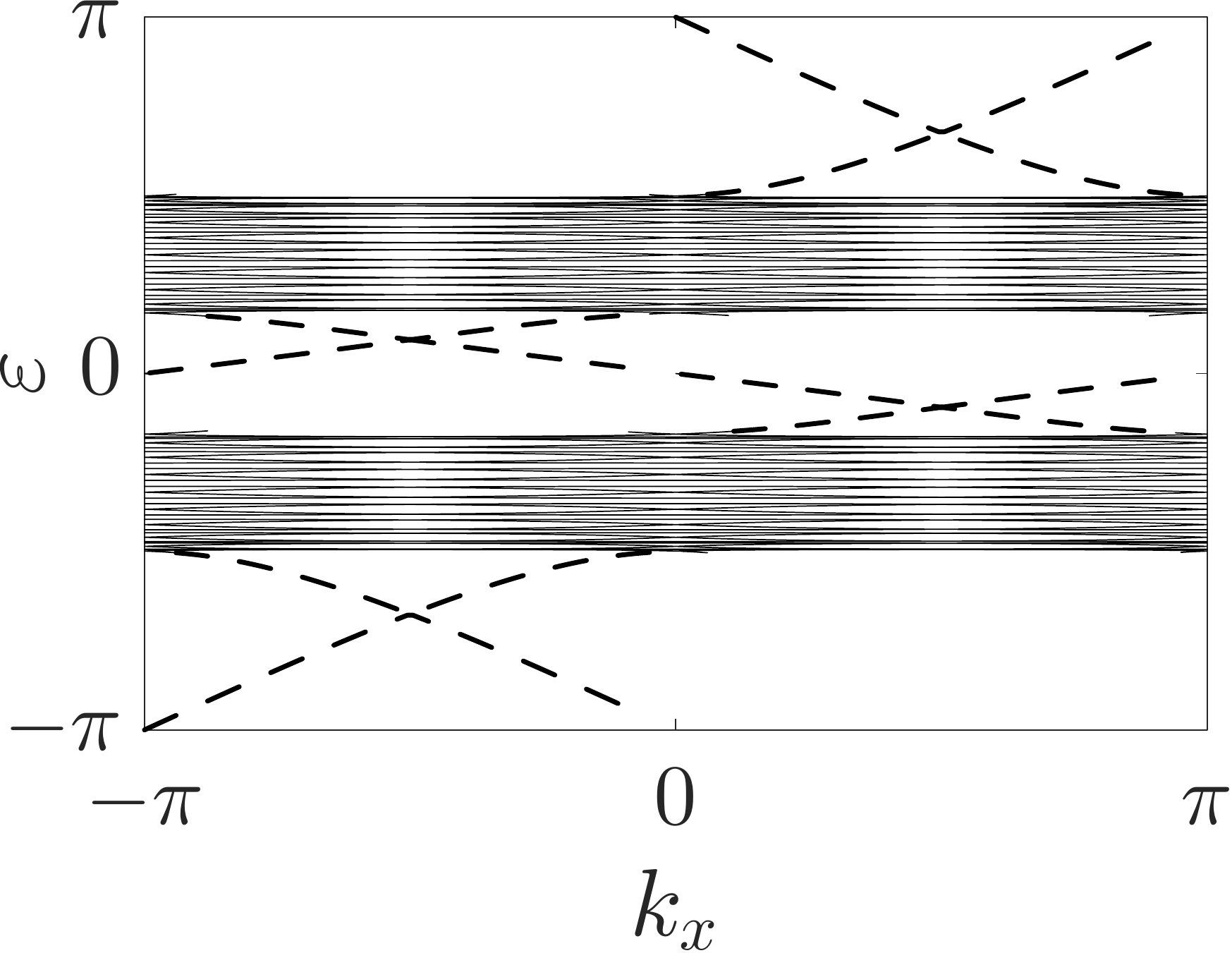}
\hspace*{\fill}
\includegraphics[scale=0.23]{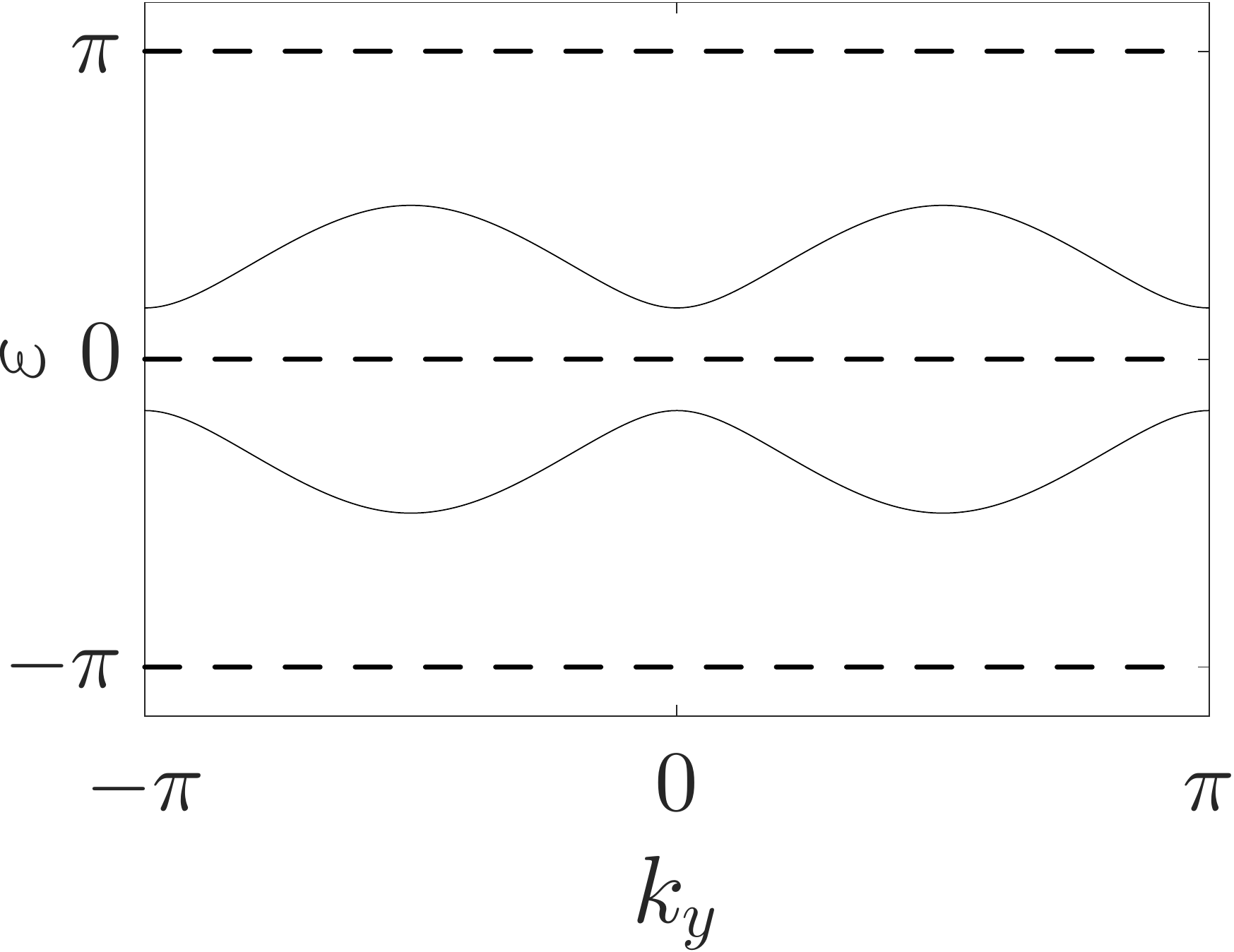}
\hspace*{\fill}
\caption{Same as Fig.~\ref{Fig:non_perfect1}, now for chiral (top row) and particle-hole symmetry (bottom row) and parameters from Table~\ref{Tab:non_perfect}.
The relevant bulk invariants are here the Chern numbers $C$ of the Floquet bands,
and the symmetry-adapted 
$W_{\mathrm{ch}}$ and $W_{\mathrm{ph}}^\alpha$-invariants from Ref.~\cite{HAF18}.
}
\label{fig:ch_ph}
\end{figure*}

\subsection{
Continuous switching between fermionic and bosonic time-reversal symmetry}
\label{sec:switch}

Since fermionic and bosonic time-reversal symmetry differ by the sign of the parameters in step 5 of the driving protocol, they are realized in separate regions of the parameter space.
Especially at perfect coupling ($J = \pm J_p$),
the conditions for fermionic or bosonic time-reversal symmetry in Table~\ref{tab:symm} are mutually exclusive.
However, considering the argument in Sec.~\ref{sec:Circumvent},
the cases $J = J_p$ and $J = - J_p$ are in fact equivalent.
The Floquet propagators in both cases differ only by a minus sign, which shifts the quasienergies by $\pi$ but affects neither the topological invariants nor the existence of boundary states.

Building on this observation, we can switch continuously between fermionic and bosonic time-reversal symmetry:
Set $J_{A \rightarrow B} = J_{C \rightarrow D} = J_p - |\delta|$ in step 2
and $J_{A \rightarrow B} = J_{C \rightarrow D} = J_p + \delta$ in step 5.
For $\delta \le 0$, the driving protocol has bosonic time-reversal symmetry.
For $\delta \ge 0$, the parameter value $J_p + \delta$ in step 5 is equivalent to the parameter value $J_p + \delta - 2 J_p = - (J_p - |\delta|)$, up to a minus sign of the Floquet propagator.
The driving protocol has fermionic time-reversal symmetry.
At perfect coupling $\delta = 0$ in steps 2 and 5, both fermionic and bosonic time-reversal symmetry are realized simultaneously.

In Fig.~\ref{fig:phase_trans} we show the change of the boundary state dispersion if the
parameter $\delta$ is varied through $\delta = 0$,
and we switch continuously from bosonic to fermionic time-reversal symmetry.
Note that since the propagator acquires a minus sign for $\delta > 0$, if compared to Fig.~\ref{Fig:non_perfect1}, the position of the gap remains at $\varepsilon = 0$.
Because of time-reversal symmetry, the boundary dispersion is invariant under the mapping $k_x \mapsto - k_x$.
While the boundary states are separated for $\delta < 0$,
they are gapless for $\delta > 0$.
Only in the latter parameter regime, the crossing at the invariant momentum $k_x = 0$ is protected by Kramers degeneracy.
In this way, continuous variation of $\delta$ 
switches between a trivial (bosonic) and non-trivial (fermionic) time-reversal symmetric topological phase, without the bulk gap closing at $\delta=0$.

\subsection{Chiral and particle-hole symmetry}

In Fig.~\ref{fig:ch_ph} we show the Floquet bands and boundary states
for chiral and particle-hole symmetry, with two gaps at quasienergies $\varepsilon=0, \pi$.
For both symmetries, we are interested in ``weak'' topological phases, where protected boundary states occur in the gap, but transport in real space is not necessarily topologically protected.

For chiral symmetry, the $W_3$-invariant has to be zero in both gaps (at $\varepsilon = 0, \pi$).
Similar to time-reversal symmetry, this implies that the boundary states are not stable under general deformations of the Floquet Hamiltonian.
However, with the alternating sign of Eq.~\eqref{eq:ch_symm}, chiral symmetry gives rise to a symmetry-protected $\mathbb Z_2$ phase that is visible in the dispersion $\varepsilon(k_{x,y})$ of boundary states in momentum space~\cite{PhysRevB.93.075405, HAF18}. 
The reason is that the dispersion fulfills the constraint $\varepsilon(k_{x,y} + \pi) \equiv - \varepsilon(k_{x,y}) \mod 2 \pi$, such that chiral symmetry protects the crossings of $\varepsilon(k_{x,y})$ through the
quasienergy $\varepsilon=0$ or $\varepsilon=\pi$. The crossings have to occur in pairs that are separated by momentum $\pi$ (see App.~B of Ref.~\cite{HAF18} for an extended argument).
Note that inclusion of the alternating sign in the symmetry relation~\eqref{eq:ch_symm} is essential for this momentum-space protection, otherwise chiral symmetry does not protect any non-trivial phase~\cite{PhysRevB.93.115429, PhysRevB.96.155118,PhysRevB.96.195303}.

In Fig.~\ref{fig:ch_ph}, exactly two crossings exist in each gap and on each boundary, which agrees with the non-zero value $W_\mathrm{ch} \ne 0$ of the $\mathbb Z_2$-valued invariant $W_\mathrm{ch}$ that is adapted to chiral symmetry~\cite{HAF18}. The above momentum-space constraint on $\varepsilon(k_{x,y})$ does not enforce that the boundary states traverse the band gap. Therefore, the chiral symmetric phase seen here does not necessarily exhibit counterpropagating boundary states with opposite chirality, and with the concomitant transport properties.

For particle-hole symmetry with $\Pi^2=1$,
topological phases are still characterized by the Chern number or, for Floquet systems, the $W_3$-invariant.
Weak topological phases, where the number of boundary states depends on the boundary orientation~\cite{PhysRevB.93.075405}, arise for vanishing $W_3$-invariant.
Several $\mathbb Z_2$-valued invariants $W_{\mathrm{ph}}^\alpha$ are required in this situation~\cite{HAF18}.
In Fig.~\ref{fig:ch_ph}, all $W_{\mathrm{ph}}^\alpha$-invariants are non-zero and boundary states exist in each gap and on each ($x$ or $y$) boundary.
Particle-hole symmetry does not enforce a zero $W_3$-invariant, but here it is $W_3 = 0$ in Fig.~\ref{fig:ch_ph}, such that the net chirality of the boundary states in each gap is zero. 

Similar to chiral symmetry, in these weak phases the appearance of boundary states in momentum space does not imply topologically protected transport in real space.
In Fig.~\ref{fig:ch_ph} the boundary state dispersion is perfectly flat on the boundary in $y$-direction while the bulk bands are dispersive (this is a particular property of the parameter set in Table~\ref{Tab:non_perfect}, not of particle-hole symmetry).
In this situation, states propagate along the $y$-direction only in the bulk but not on the boundary.

 In contrast to time-reversal symmetry, which requires the Kane-Mele invariant of the Floquet bands, the Chern number remains a relevant invariant for chiral and particle-hole symmetry. In Fig.~\ref{fig:ch_ph} the Chern numbers of all Floquet bands are zero. Therefore, the boundary states observed here belong to anomalous Floquet topological phases,
 and appear although the individual Floquet bands are topologically trivial.

\subsection{Propagation of boundary states}

\begin{figure}
\hspace*{\fill}
\includegraphics[scale=1]{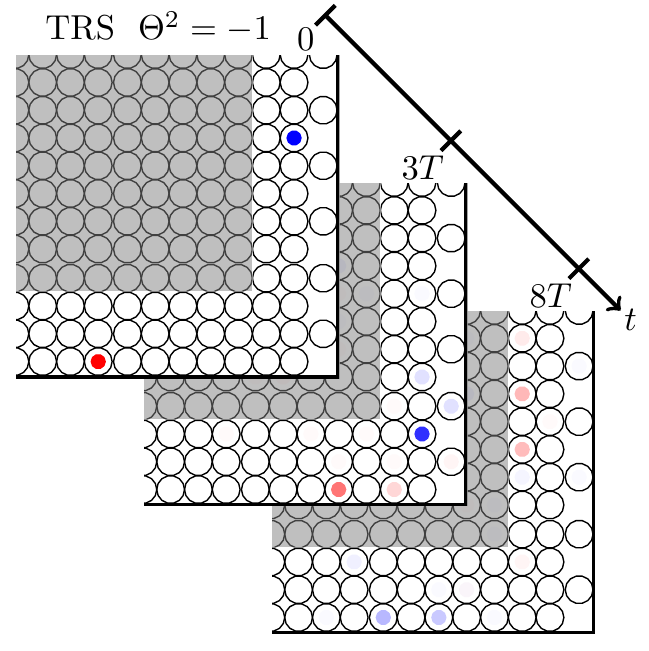}
\hspace*{\fill}
\\
\hspace*{\fill}
\includegraphics[scale=1]{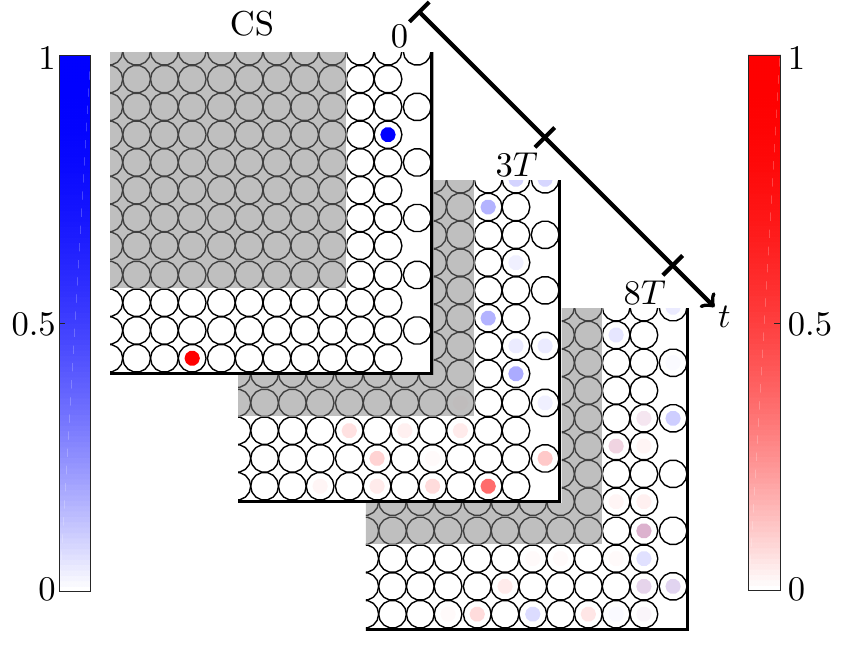}
\hspace*{\fill}
\\
\hspace*{\fill}
\includegraphics[scale=1]{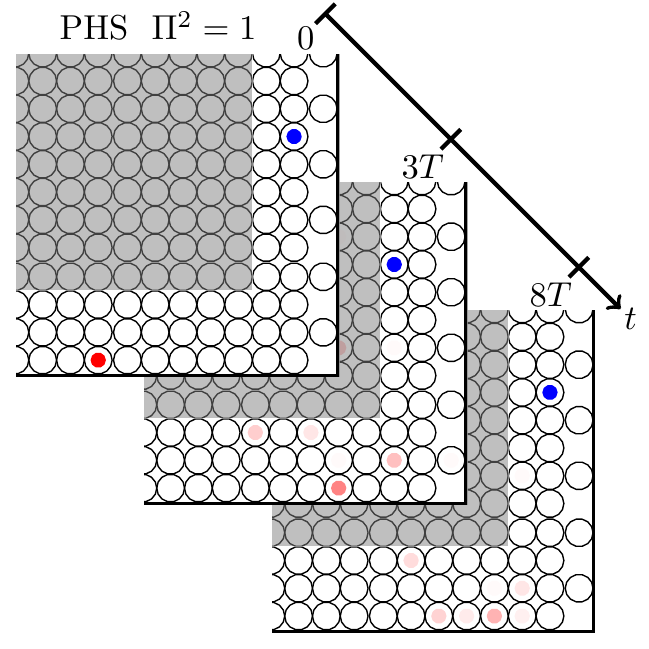}
\hspace*{\fill}
\caption{Propagation of boundary states in the vicinity of a corner,
starting from a ``red''  $A$ site or a ``blue'' $B$ site.
Open black circles indicate the lattice sites. Shown is the (squared) wave function amplitude, with colors according to the two color bars, after zero ($t=0$), three ($t=3T$), or eight cycles ($t=8T$) of driving protocol \textsf{A}. 
}
\label{Fig:edge_state_prop}
\end{figure}

In Fig.~\ref{Fig:edge_state_prop} we show the real-space propagation of boundary states in the vicinity of a corner.
At $t=0$, an initial state is prepared either on a ``red'' $A$ site  of the horizontal boundary in the $x$-direction or on a ``blue''  $B$ site of the vertical boundary in the $y$-direction, and then observed after three ($t = 3T$) and eight ($t = 8 T$) periods of the driving protocol \textsf{A}.

Since the parameter values in Table~\ref{Tab:non_perfect} are sufficiently close to perfect coupling such that the essential patterns of motion from Fig.~\ref{Fig:sketch_perfect} still survive,
the ``red'' (or ``blue'') state propagates mainly counterclockwise (or clockwise).
Note that the amplitude at the boundary decreases over time since the state propagates partially into the bulk.
Also, since the boundary state dispersion is not perfectly linear (see Figs.~\ref{Fig:non_perfect1},~\ref{fig:ch_ph}), the state is distributed over several lattice sites at later propagation times.

As soon as the state hits the corner, it either propagates around the corner without backscattering (for fermionic time-reversal symmetry), or is partially (for chiral symmetry) or totally (for particle-hole symmetry) reflected.
This behavior can be attributed to the different nature of the (weak) topological phases for the different symmetries:
For fermionic time-reversal symmetry, transport is topologically protected.
For chiral symmetry, the boundary states are still protected in momentum space but the dispersion along the $y$-boundary does not traverse the band gap, which leads to partial reflection.
For particle-hole symmetry, the boundary state dispersion along the $y$-direction is perfectly flat,
which leads to total reflection of states starting on the $x$-boundary.
States on the $y$-boundary stay within one unit cell, moving back and forth between the initial $B$ site and the adjacent $A$ site with each period of the driving protocol.

\begin{figure*}
\hspace*{\fill}
\includegraphics[scale=0.8]{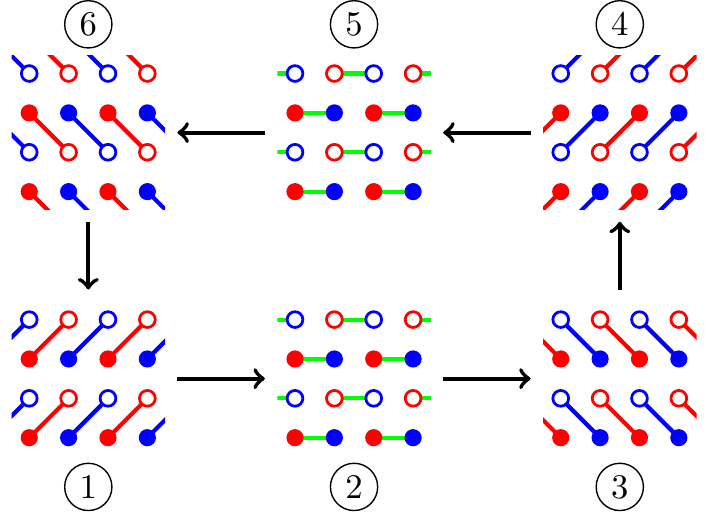}
\hspace*{\fill}
\includegraphics[scale=0.8]{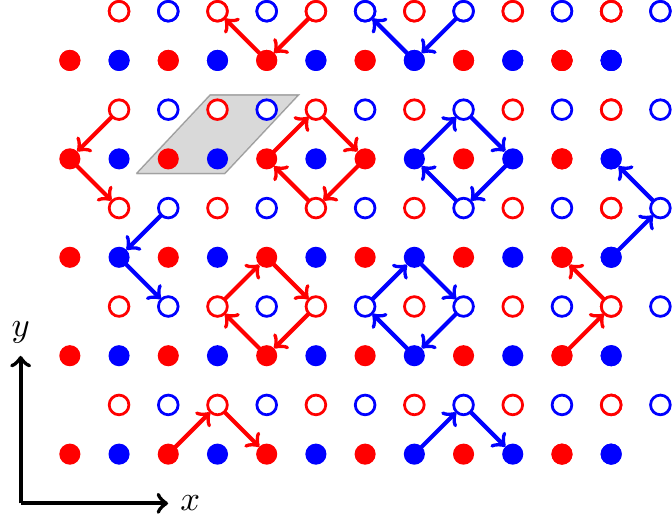}
\hspace*{\fill}
\includegraphics[scale=0.25]{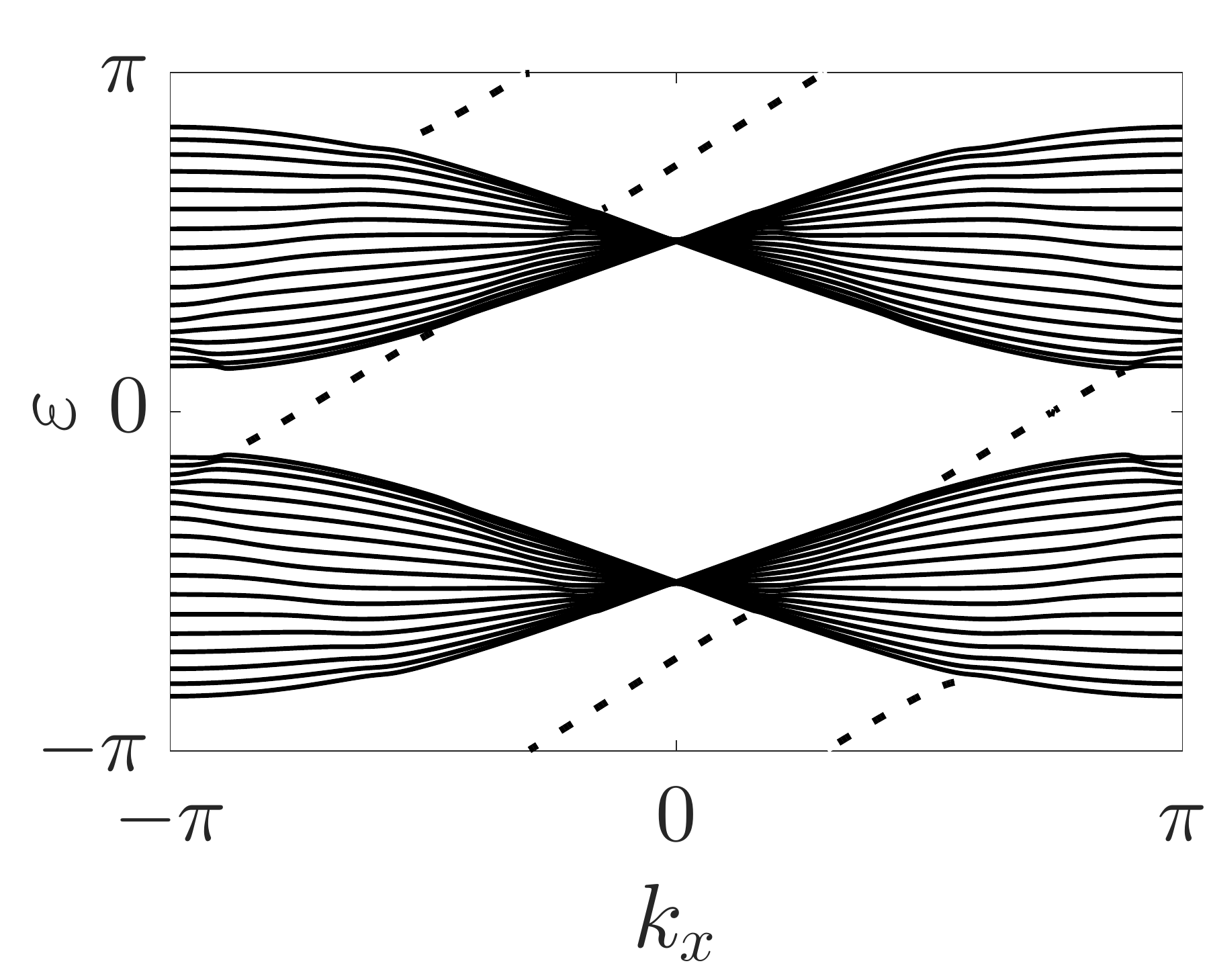}
\hspace*{\fill}
\caption{Left panel: Six-step driving protocol \textsf{B} for particle-hole symmetry $\Pi^2=-1$.
Central panel: Patterns of motion during one cycle at perfect coupling.
Right panel: Floquet bands and boundary states for the set of parameters from Tab.~\ref{Tab:PH}.}
\label{Fig:PH}
\end{figure*}

\section{Universal driving protocol \textsf{B}: Particle-hole symmetry} 
\label{sec:PH}

For particle-hole symmetry with $\Pi^2 = -1$, again the symmetry operator $S_8$ has to be used for construction of the driving protocol.
Now, the symmetry relation~\eqref{eq:ph_symm} contains the same time argument on both sides,
and according to Sec.~\ref{sec:principal} we have to use the parallel diagonal coupling patterns (f1)--(f4) from Fig.~\ref{fig:para_diag} in App.~\ref{app:Parallel}.

Repetition of the procedure from Sec.~\ref{Sec_4}
leads to the driving protocol \textsf{B} in Fig.~\ref{Fig:PH}.
The considerations from Sec.~\ref{sec:EquivalentProtocols} can be adapted to construct two variants of the protocol,
and the strategy from Sec.~\ref{sec:Circumvent} allows for replacement of negative by positive couplings.

The patterns of motion for perfect coupling ($J_{s \circ s'} = J_p$ in steps 1,3,4,6 and $J_{s \circ s'} = 0$ in steps 2,5) are shown in the central panel of Fig.~\ref{Fig:PH}. Comparison with Fig.~\ref{Fig:sketch_perfect} shows that now states on the ``red'' and ``blue'' sublattice propagate in the same direction.
This explains, quite intuitively, why parallel (perpendicular) diagonal coupling patterns are used for particle-hole (time-reversal) symmetry with copropagating (counterpropagating) boundary states.

For the general case, we use the parameter values in Table~\ref{Tab:PH}.
The corresponding Floquet bands and boundary states are shown in the right panel of Fig.~\ref{Fig:PH}. Two boundary states with the same chirality exist in the two gaps at quasienergies $\varepsilon=0$ and $\varepsilon=\pi$. This phase is characterized by the conventional Chern number $C$ and $W_3$-invariant, which are restricted to even values ($2 \mathbb Z$) by the particle-hole symmetry. In accordance with the appearance of two copropagating boundary states, we have $W_3=2$ for both gaps.
Consequently, we have $C=0$ for the individual Floquet bands, which is the signature of an anomalous Floquet topological phase with $C=0$ but $W_3 \ne 0$.

\begin{table}
\caption{Similar to Tab.~\ref{Tab:non_perfect}, parameter set for driving protocol \textsf{B} with particle-hole symmetry $\Pi^2=-1$.
In Fig.~\ref{Fig:PH} we use the values of the free parameters $\Delta, J, J'$ specified under ``this work''.
}
\begin{center}
\begin{ruledtabular}
\begin{tabular}{llll}
 &  $\mathrm{PHS}\; \Pi^2=-1$
\\ \hline
 step $1$ &  $J_{A\nearrow C}=\phantom{-}J_{\mathrm{p}}$  &  step $4$ & $J_{C\nearrow A}=\phantom{-}J_{\mathrm{p}}$\\
&  $J_{B\nearrow D}=-J_{\mathrm{p}}$ &  & $J_{D\nearrow B}=-J_{\mathrm{p}}$\\
 & $\Delta_{A}=\Delta_{D}= \phantom{-} \Delta$  & & $\Delta_{A}=\Delta_{D}= \phantom{-} \Delta$\\
 & $\Delta_{B}=\Delta_{C}= - \Delta$  & & $\Delta_{B}=\Delta_{C}= - \Delta$\\[3pt]  \cline{2-2} \cline{4-4}
 
step $2$ & $J_{A\rightarrow B}=J$ & step $5$ & $J_{A\rightarrow B}=J'$\\
& $J_{C\rightarrow D}=J$ & & $J_{C\rightarrow D}=J'$ \\[3pt] \cline{2-2} \cline{4-4}

 step $3$ &  $J_{C\searrow A}=\phantom{-}J_{\mathrm{p}}$  & step $6$ & $J_{A\searrow C}=\phantom{-}J_{\mathrm{p}}$\\
&  $J_{D \searrow B}=-J_{\mathrm{p}}$ & & $J_{B\searrow D}=-J_{\mathrm{p}}$  \\
 & $\Delta_{A}=\Delta_{D}= \phantom{-} \Delta$  & & $\Delta_{A}=\Delta_{D}= \phantom{-} \Delta$\\
 & $\Delta_{B}=\Delta_{C}= - \Delta$  & & $\Delta_{B}=\Delta_{C}= - \Delta$ 

 \\[3pt] \hline
this &  $J = 2 \pi/T$ & & $J'=\pi/T$  \\
work &  $\Delta = 3/T$ && 
\end{tabular}
\end{ruledtabular}
\end{center}
\label{Tab:PH}
\end{table}

\section{Conclusions}
\label{Sec_6}

The universal driving protocol introduced in the present paper allows for the realization of Floquet topological phases with time-reversal, chiral, or particle-hole symmetry.
Switching between the different symmetries only requires adjustment of a few parameters, or
the replacement of parallel (protocol \textsf{B}) with perpendicular (protocol \textsf{A}) diagonal couplings.
The general structure of the driving protocol,
which follows from the analysis of the possible symmetry operators for the underlying square lattice Hamiltonian,
remains unchanged.
In fact, if we allow for coupling of three or more lattice sites, the two types \textsf{A} and \textsf{B} of the universal driving protocol are continuously connected, and appear as special cases of the slightly generalized universal driving model depicted in App.~\ref{app:Combined}.

Due to the minimal complexity of the universal driving protocol,
which is a result of the constraints accounted for in its construction, 
it is not only of theoretical value but can be implemented by extension of previous experimental work~\cite{anomalous, anomalous_2}.
Ref.~\cite{Maczewsky_et_al} documents the photonic lattice implementation of the driving protocol with fermionic time-reversal symmetry, and reports the observation of a topological phase with scatter-free counterpropagating boundary states. These states are protected by the fermionic time-reversal symmetry prescribed by the protocol, even though the underlying photonic system is of bosonic nature.

A novel aspect yet to be explored in more detail is the possibility of switching
between fermionic and bosonic time-reversal symmetry by continuous variation of a parameter.
Normally, without symmetries, switching between non-trivial and trivial topological phases requires that a gap closes and reopens.
The driving protocol allows us to switch between a non-trivial and trivial symmetry-protected topological phase
without directly affecting the topological nature of the system (the gap stays open), and without breaking time-reversal symmetry. Instead, only the type of time-reversal symmetry changes,
and that even in a continuous manner.

\begin{acknowledgments}
The authors would like to thank L. Maczewsky, M. Kremer, A. Szameit, and A. Fritzsche for useful discussions on time-reversal symmetric Floquet insulators.
\end{acknowledgments}

\appendix

\begin{figure}[b]
\hspace*{\fill}
\includegraphics[width=0.45\linewidth]{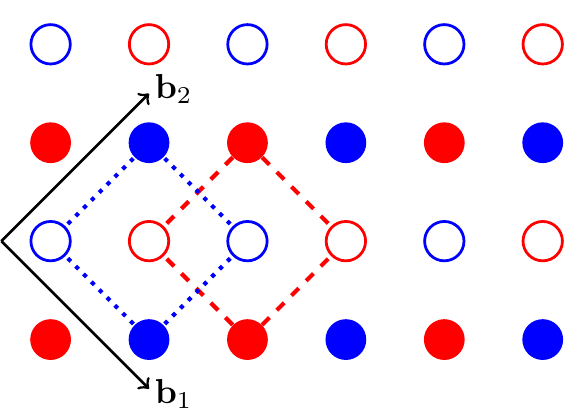}
\hspace*{\fill}
\caption{The square lattice can be viewed as a centered square lattice, or the union of a ``red'' and ``blue'' square lattice.}
\label{fig:CenteredSquare}
\end{figure}

\begin{figure*}
\hspace*{\fill}
\includegraphics[scale=0.9]{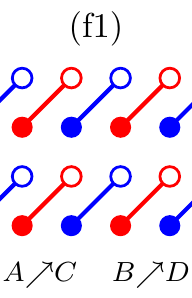}
\hspace*{\fill}
\includegraphics[scale=0.9]{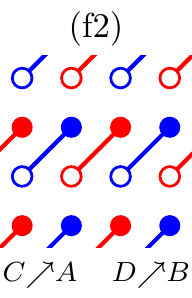}
\hspace*{\fill}
\includegraphics[scale=0.9]{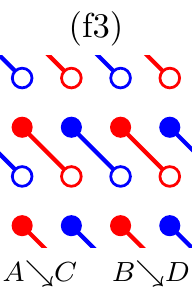}
\hspace*{\fill}
\includegraphics[scale=0.9]{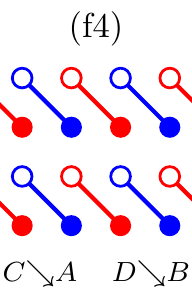}
\hspace*{\fill}
\includegraphics[scale=0.9]{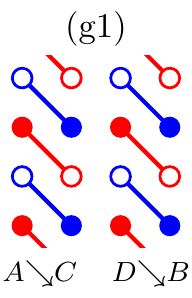}
\hspace*{\fill}
\includegraphics[scale=0.9]{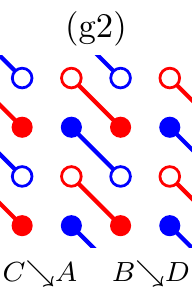}
\hspace*{\fill}
\includegraphics[scale=0.9]{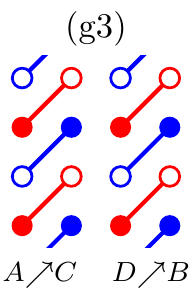}
\hspace*{\fill}
\includegraphics[scale=0.9]{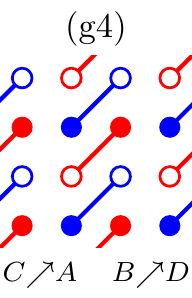}
\hspace*{\fill}
\caption{The eight coupling patterns with parallel diagonal pairwise couplings.
Patterns (f1) and (g1) correspond to patterns (f) and (g) in Fig.~\ref{fig:Lattice}.
}
\label{fig:para_diag}
\end{figure*}

\section{Pseudo-spin interpretation}
\label{app:Pseudo}

The pseudo-spin interpretation of the ``red'' and ``blue'' sublattice structure depicted in Figs.~\ref{fig:Lattice}--\ref{Fig:sketch_perfect} is suggested by the geometric structure of the operator $S_8$ in Fig.~\ref{fig:Symm_op}. A natural way to represent the pseudo-spin is to understand the original lattice as a centered square lattice (see Fig.~\ref{fig:CenteredSquare}), and associate the ``red'' (``blue'') sublattice with the ``up'' (``down'') component of a spin $\tfrac12$.

Technically, the pseudo-spin interpretation is obtained through a Hilbert space isomorphism $\mathcal {I_S}$, which is defined by the mapping
\begin{equation}
\begin{split}
\mathcal{I_S} \, |i \vec b_1 +  j \vec b_2 \rangle \otimes |{\uparrow}\rangle \;  &= \; |i \vec b_1 + j \vec b_2 \rangle \;, \\
\mathcal{I_S} \,     |i \vec b_1 +  j \vec b_2 \rangle \otimes |{\downarrow}\rangle \; &= \; |i \vec b_1 + j \vec b_2 + \vec e_x \rangle \;,
    \end{split}
\end{equation}
for $i,j \in \mathbb Z$.
Here, $\vec b_{1}=(1,-1)^t, \vec b_2=(1,1)^t$ are the translation vectors of the centered square lattice,
and $\vec e_{x}=(1,0)^t, \vec e_y=(0,1)^t$  the unit vectors of the original square lattice.
 In terms of the vectors $\vec a_x, \vec a_y, \boldsymbol\delta_s$ used in Sec.~\ref{Sec_2}, we have
\begin{equation}
\mathcal{I_S} \,  |i \vec b_1 +  j \vec b_2 \rangle \otimes |\mathcal S\rangle \;  = \; | \lfloor\tfrac{i+j}{2}\rfloor \vec a_x + \lfloor\tfrac{j-i}{2}\rfloor \vec a_y + \boldsymbol \delta_s \rangle \;,
\end{equation}
where $\lfloor \cdot \rfloor$ denotes the floor function (rounding down to the next integer),
and $s$ is chosen according to
\begin{equation}
\begin{array}{l|cc}
& \; \mathcal S = \uparrow \; & \; \mathcal S = \downarrow \; \\\cline{1-3}
\rule{0pt}{13pt} i+j \text{ even } & s = A  & s = B \\
\rule{0pt}{13pt} i+j \text{ odd } & s = C & s = D
\end{array} \raisebox{-18pt}{\;\;\;.}
\end{equation}

Within the pseudo-spin interpretation, diagonal pairwise couplings correspond to translations along the vectors $\vec b_1, \vec b_2$ that preserve the pseudo-spin, as in
\begin{equation}
\begin{split}
(\mathcal {I_S}^{-1} \hat t_{A \nearrow C  } \mathcal {I_S}) \, |i \vec b_1 +  j \vec b_2 \rangle \otimes |{\uparrow}\rangle  & =|(i+1) \vec b_1 +  j \vec b_2 \rangle \otimes |{\uparrow}\rangle  \;, \\
(\mathcal {I_S}^{-1} \hat t_{A \nearrow C  } \mathcal {I_S}) \, |i \vec b_1 +  j \vec b_2 \rangle \otimes |{\downarrow}\rangle  & =0 \;.
 \end{split}
\end{equation}
The horizontal coupling pattern (a) in Fig.~\ref{fig:Lattice}, which appears in steps 2 and 5 of the universal driving protocol, corresponds to a spin transformation 
\begin{equation}
\mathcal {I_S}^{-1} ( \hat t_{A \rightarrow B} + \hat t_{A \rightarrow B}^\dagger + \hat t_ {C \rightarrow D} + \hat t_ {C \rightarrow D}^\dagger) \mathcal{I_S} = \sigma_x
\end{equation}
with the Pauli matrix $\sigma_x$ that preserves the $i, j$ index of the centered square lattice.
Note that here the parameters $J_{A \rightarrow B}$, $J_{C \rightarrow D}$ of the two pairwise couplings are equal (cf. Table~\ref{Tab:non_perfect}).
The remaining horizontal and vertical couplings, which are not compatible with the symmetry operator $S_8$, have no such simple representation.

The operator $S_8$ itself allows for a simple representation if the matrices $\sigma$, $\tau$ in Eq.~\eqref{S18} are given by a common $2 \times 2$ matrix $\Sigma$, i.e.,
$\sigma = \tau = \Sigma$.
Then, we simple have
\begin{equation}
 \mathcal {I_S}^{-1} \, S_8 \, \mathcal {I_S}= \Sigma \;.
\end{equation}
At least for fermionic time-reversal symmetry, this form of $S_8$ is mandatory (with $\Sigma = \sigma_y$), up to phase factors in $\sigma$, $\tau$.
Note that these phase factors could be absorbed into the mapping $\mathcal{I_S}$, preserving the simple form of $S_8$ even in the general case.

The pseudo-spin interpretation of the square lattice allows us to reuse familiar notions such as ``helicity'' of boundary states in the present context.
Conversely, the existence of this interpretation, as well as the precise form of the mapping $\mathcal{I_S}$ of the (pseudo-) spin onto the square lattice,
is a natural consequence of the symmetry analysis provided in the present paper.

\section{Parallel diagonal couplings}
\label{app:Parallel}

On the square lattice with a four-element unit cell, $4 \times 4 = 16$ diagonal coupling patterns exist in total.
Four of them contain pairwise couplings that cross each other, and are not allowed due to the constraints imposed in Sec.~\ref{Sec_2}.
Out of the allowed twelve patterns,
the four perpendicular diagonal coupling patterns (b)--(e) in Fig.~\ref{fig:Lattice} constitute the main steps of the driving protocol \textsf{A} with time-reversal symmetry from Sec.~\ref{Sec_4}.
Out of the remaining eight parallel diagonal coupling patterns depicted in Fig.~\ref{fig:para_diag},
patterns (f1)--(f4) constitute the main steps of the driving protocol \textsf{B} with particle-hole symmetry in Sec.~\ref{sec:PH}.
The latter choice is mandatory, because only these patterns are mapped onto themselves by the symmetry operator $S_8$, while patterns (g1)$\leftrightarrow$(g2) and (g3)$\leftrightarrow$(g4) are swapped.

That leaves open the question why the parallel diagonal coupling patterns are not used for the driving protocol \textsf{A} with time-reversal symmetry.
Intuitively, this question is answered by comparison of the patterns of motion in Fig.~\ref{Fig:sketch_perfect} and Fig.~\ref{Fig:PH}: Parallel diagonal couplings give rise to copropagating states, while time-reversal symmetry requires counterpropagating states, hence the perpendicular diagonal coupling patterns.

For a more exhaustive argument, consider the situation that the driving protocol should support time-reversal symmetry,
but  has to be composed only out of parallel diagonal coupling patterns.
We can then try to repeat the construction from Sec.~\ref{sec:mini:a} and focus on the two central steps, e.g., steps 3,~4 in a six-step protocol. These two steps must be exchanged under a mapping with $S_8$.

\begin{figure}[b]
\hspace*{\fill}
\includegraphics[width=0.8\linewidth]{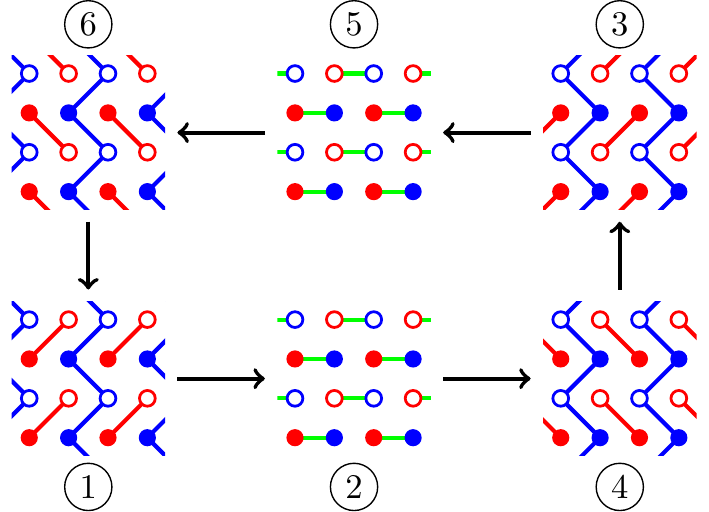}
\hspace*{\fill}
\caption{A universal driving protocol that contains protocol~\textsf{A} (right variant in Fig.~\ref{Fig:six_step_protocol}) and protocol~\textsf{B},
but violates the constraints from Sec.~\ref{Sec_2}.}
\label{fig:combined_proto}
\end{figure}

If the two steps involve patterns (f1)--(f4), they are mapped onto each other by $S_8$, and can be combined into a single step. 
In this way, nothing is gained for the construction of the driving protocol.
If the two steps involve patterns (g1)--(g4), possible combinations are pattern (g1) followed by pattern (g2), or patterns (g3) followed by pattern (g4), etc. 
Visual inspection of these patterns in Fig.~\ref{fig:para_diag} shows that such combinations transport states by two lattice sites in diagonal direction, but not on a closed loop as required for our driving protocol (see Fig.~\ref{Fig:sketch_perfect}).
Nothing is gained for the construction of the driving protocol in this way, either.

We conclude that a driving protocol with time-reversal symmetry has to use the perpendicular diagonal coupling patterns (b)--(e) from Fig.~\ref{fig:Lattice}, instead of
the parallel diagonal coupling patterns (f1)--(g4) from Fig.~\ref{fig:para_diag}.

\section{Joint \textsf{A} and \textsf{B} driving protocol}
\label{app:Combined}

In the main text, the two types \textsf{A} (in Fig.~\ref{Fig:six_step_protocol}) and \textsf{B} (in Fig.~\ref{Fig:PH}) of the driving protocol appear as disjoint cases, with either perpendicular or parallel diagonal couplings.
In fact, both types of the protocol are just special cases of the combined driving protocol shown in Fig.~\ref{fig:combined_proto}.
However, continuous interpolation between protocol \textsf{A} and protocol \textsf{B} requires inclusion of couplings between three or more lattice sites, as is evident from the zigzag ``blue'' couplings in Fig.~\ref{fig:combined_proto}.
The inclusion of such couplings is perfectly valid, unless we impose the very restrictive constraints of Sec.~\ref{Sec_2}.
Only because of these constraints, we had to discuss protocol \textsf{A} and protocol \textsf{B} separately in the main text.

\section{Protocols with a two-site unit cell}
\label{app:nogo}

\begin{figure}[b]
\hspace*{\fill}
\includegraphics[scale=0.9]{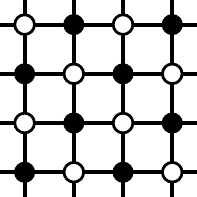}
\hspace*{\fill}
\includegraphics[scale=0.9]{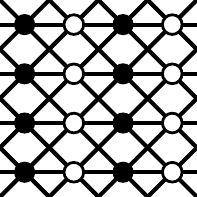}
\hspace*{\fill}
\caption{Pairwise couplings (solid lines) on a square lattice with a two-element (a filled and an open circle) unit cell, and two choices for translation symmetry.}
\label{Fig:model_two}
\end{figure}

Translational symmetry on a square lattice with a two-element unit cell can be implemented in two ways (see Fig.~\ref{Fig:model_two}):
either with primitive translation vectors $\vec a_x = (2,0)$, $\vec a_y = (1,1)$ (left panel), or
 $\vec a_x = (2,0)$, $\vec a_y = (1,0)$ (right panel).
With the constraint that pairwise couplings are allowed only between neighboring lattice sites,
only the couplings depicted in Fig.~\ref{Fig:model_two} are possible.
In the two cases, either (anti-)diagonal (left panel) or vertical (right panel) coupling terms are forbidden.

\begin{figure}[b]
\hspace*{\fill}
\includegraphics[scale=0.9]{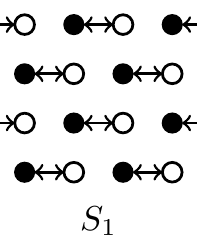}
\hspace*{\fill}
\includegraphics[scale=0.9]{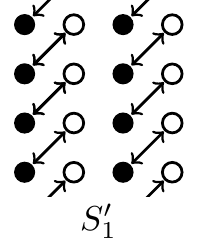}
\hspace*{\fill}
\includegraphics[scale=0.9]{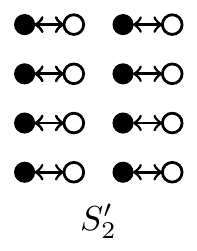}
\hspace*{\fill}
\\
\hspace*{\fill}
\includegraphics[scale=0.9]{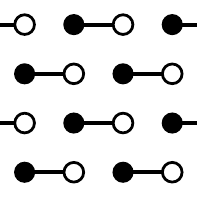}
\hspace*{\fill}
\includegraphics[scale=0.9]{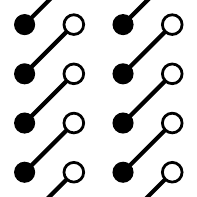}
\hspace*{\fill}
\includegraphics[scale=0.9]{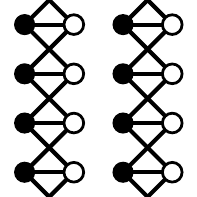}
\hspace*{\fill}
\caption{Options for a symmetry operator for fermionic time-reversal symmetry on the square lattice from Fig.~\ref{Fig:model_two} (top row), and the compatible pairwise couplings (bottom row).}
\label{fig:Symm_op_two}
\end{figure}

A symmetry analysis in the spirit of Sec.~\ref{sec:symm} leaves us with only three options
for a symmetry operator that could be used to implement fermionic time-reversal symmetry  (see Fig.~\ref{fig:Symm_op_two}).
For all options, the pairwise couplings compatible with the symmetry do not connect the entire lattice.
We conclude that, under the constraints imposed here, a non-trivial $2+1$-dimensional topological phase with time-reversal symmetry cannot be realized with a two-element unit cell, but requires at least a four-element unit cell.

\end{document}